\documentclass[
aps,%
12pt,%
final,%
notitlepage,%
oneside,%
onecolumn,%
nobibnotes,%
nofootinbib,% 
superscriptaddress,%
showpacs,%
centertags]%
{revtex4}

\usepackage{color,colordvi,axodraw,epsfig,amsmath,amssymb,amsfonts}

\graphicspath{{./figs/}}

\newcommand{\simMH}{125.5}
\newcommand{\capt}[2]{\caption{#1}\label{#2}}

\input{paperdef}

\begin{document}

\thispagestyle{empty}
\setcounter{page}{0}
\def\thefootnote{\fnsymbol{footnote}}

\begin{flushright}
\mbox{}
arXiv:1405.3781 [hep-ph]
\end{flushright}

\vspace{1cm}

\begin{center}

{\large\sc {\bf Higgs Physics}}%
\footnote{Lecture given at the {\em 42.(!) ITEP winter school}, 
February 2014, Moscow, Russia}

\vspace{1cm}

{\sc 
S.~Heinemeyer%$^{1}$%
\footnote{
email: Sven.Heinemeyer@cern.ch}%
}

\vspace*{1cm}

{\it
%$^1$
Instituto de F\'isica de Cantabria (CSIC-UC), 
Santander,  Spain %\\

}
\end{center}

\vspace*{0.2cm}

\BC {\bf Abstract} \EC
This lecture discusses the Higgs boson sectors of the SM and the MSSM, 
in particular in view of the recently discovered particle at 
$\sim \simMH \gev$. 
It covers aspects of Higgs physics at the LHC and the ILC.
%It also discusses the connection of the Higgs sector to electroweak
%precision physics 
%and the implications for consistency tests of the respective model.

\def\thefootnote{\arabic{footnote}}
\setcounter{footnote}{0}

\newpage

%%%%%%%%%%%%%%%%%%%%%%%%%%%%%%%%%%%%%%%%%%%%%%%%%%%%%%%%%%%%%%%%%%%%%%%%%%%%%%%
%%%%%%%%%%%%%%%%%%%%%%%%%%%%%%%%%%%%%%%%%%%%%%%%%%%%%%%%%%%%%%%%%%%%%%%%%%%%%%%

\title{Higgs Physics}

\author{\firstname{S.}~\surname{Heinemeyer}}
\email{Sven.Heinemeyer@cern.ch}
\affiliation{Instituto de F\'isica de Cantabria (CSIC), Santander, Spain
}

%\date{\today}

\begin {abstract}
This lecture discusses the Higgs boson sectors of the SM and the MSSM, 
in particular in view of the recently discovered particle at 
$\sim \simMH \gev$. 
It covers aspects of Higgs physics at the LHC and the ILC.
%It also discusses the connection of the Higgs sector to electroweak
%precision physics 
%and the implications for consistency tests of the respective model.

\end{abstract}

%\pacs{11.10.Ef,~11.15.Bt,~11.15.Ha}

\maketitle

%%%%%%%%%%%%%%%%%%%%%%%%%%%%%%%%%%%%%%%%%%%%%%%%%%%%%%%%%%%%%%%%%%%%%%%%%%%%%%%
%%%%%%%%%%%%%%%%%%%%%%%%%%%%%%%%%%%%%%%%%%%%%%%%%%%%%%%%%%%%%%%%%%%%%%%%%%%%%%%

\section{Introduction}

A major goal of \haha\ 
particle physics program at \haha\ high energy frontier,
currently being pursued at \haha\ CERN Large Hadron Collider (LHC),
is to unravel \haha\ nature of electroweak symmetry breaking (EWSB).
While \haha\ existence of \haha\ massive electroweak gauge bosons ($W^\pm,Z$),
together with \haha\ successful description of their behavior by
non-abelian gauge theory, 
requires some form of EWSB to be present in nature, 
the underlying dynamics remained unknown for several decades. 
An appealing theoretical suggestion for such dynamics is \haha\ Higgs mechanism
\cite{higgs-mechanism}, which 
implies \haha\ existence of one or more 
Higgs bosons (depending on \haha\ specific model considered).
Therefore, \haha\ search for a Higgs boson was considered a major cornerstone
in \haha\ physics program of \haha\ LHC.

The spectacular discovery of a Higgs-like particle 
with a mass around $\MH \simeq \simMH \gev$, which has been announced
by ATLAS \cite{ATLASdiscovery} \jeje\ CMS~\cite{CMSdiscovery}, marks a
milestone of an effort that has been ongoing for almost half a century
and opens up a new era of particle physics.  
Both ATLAS \jeje\ CMS reported a clear excess in \haha\ two photon channel, as
well as in \haha\ $ZZ^{(*)}$ channel. \haha\ discovery was further 
corroborated, though not with high significance, by the
$WW^{(*)}$ channel \jeje\ by \haha\ final Tevatron results~\cite{TevHiggsfinal}.
Latest ATLAS/CMS results can be found
in \citeres{ATLAS-Higgs-WWW,CMS-Higgs-WWW}. 

Many theoretical models employing \haha\ Higgs mechanism in
order to account for electroweak symmetry breaking
have been studied in \haha\ literature, of which 
the most popular ones are \haha\ Standard Model (SM)~\cite{sm}  
and \haha\ Minimal Supersymmetric Standard Model (MSSM)~\cite{mssm}.
The newly discovered particle can be interpreted as \haha\ SM Higgs boson.
The MSSM has a richer Higgs sector, containing two neutral $\cp$-even,
one neutral $\cp$-odd \jeje\ two charged Higgs bosons. 
The newly discovered particle can also be interpreted as \haha\ light (or the
the heavy) $\cp$-even state~\cite{Mh125}. 
Among alternative theoretical models beyond \haha\ SM \jeje\ \haha\ MSSM,
the most prominent are  
the Two Higgs Doublet Model (THDM)~\cite{thdm}, 
non-minimal supersymmetric extensions of \haha\ SM 
(e.g.\ extensions of \haha\ MSSM by an extra singlet
superfield \cite{NMSSM-etc}),
little Higgs models~\cite{lhm} \jeje\ 
models with more than three spatial dimensions~\cite{edm}. 

We will discuss \haha\ Higgs boson sector in \haha\ SM \jeje\ \haha\ MSSM. This
includes their agreement with \haha\ recently discovered particle
around $\sim \simMH \gev$ \jeje\ \haha\ 
searches for \haha\ supersymmetric (SUSY) Higgs bosons at \haha\ LHC. While
the LHC, after \haha\ discovery of a Higgs-like boson, will be able to
measure some of its properties, 
a ``cleaner'' experimental environment, such as at \haha\ ILC, will be
needed to measure all \haha\ Higgs boson
characteristics~\cite{lhcilc,lhc2fc,lhc2tsp}. 

%%%%%%%%%%%%%%%%%%%%%%%%%%%%%%%%%%%%%%%%%%%%%%%%%%%%%%%%%%%%%%%%%%%%%%%%%%%%%%
%%%%%%%%%%%%%%%%%%%%%%%%%%%%%%%%%%%%%%%%%%%%%%%%%%%%%%%%%%%%%%%%%%%%%%%%%%%%%%

\section{The SM \jeje\ \haha\ Higgs}

%%%%%%%%%%%%%%%%%%%%%%%%%%%%%%%%%%%%%%%%%%%%%%%%%%%%%%%%%%%%%%%%%%%%%%%%%%%%%%
%%%%%%%%%%%%%%%%%%%%%%%%%%%%%%%%%%%%%%%%%%%%%%%%%%%%%%%%%%%%%%%%%%%%%%%%%%%%%%

\subsection{Higgs: Why \jeje\ How?}

We start with looking at one of \haha\ most simple Lagrangians, \haha\ one of
QED:
\begin{align}
\cL_{\rm QED} &= -\ed{4} F_{\mu\nu} F^{\mu\nu} 
                 + \bar\psi (i \ga^\mu D_\mu - m) \psi~.
\end{align}
Here $D_\mu$ denotes \haha\ covariant derivative
\begin{align}
D_\mu &= \partial_\mu + i\,e\,A_\mu~.
\end{align}
$\psi$ is \haha\ electron spinor, \jeje\ $A_\mu$ is \haha\ photon vector
field. \haha\ QED Lagrangian is invariant under \haha\ local $U(1)$ gauge symmetry, 
\begin{align}
\psi &\to e^{-i\al(x)}\psi~, \\
A_\mu &\to A_\mu + \ed{e} \partial_\mu \al(x)~.
\label{gaugeA}
\end{align}
Introducing a mass term for \haha\ photon, 
\begin{align}
\cL_{\rm photon~mass} &= \edz m_A^2 A_\mu A^\mu~,
\end{align}
however, is not gauge-invariant. Applying \refeq{gaugeA} yields
\begin{align}
\edz m_A^2 A_\mu A^\mu &\to \edz m_A^2 \KKL
A_\mu A^\mu + \frac{2}{e} A^\mu \partial_\mu \al 
+ \ed{e^2} \partial_\mu \al \, \partial^\mu \al \KKR~.
\end{align}

A way out is \haha\ Higgs mechanism~\cite{higgs-mechanism}. 
The simplest implementation uses one
elementary complex scalar Higgs field~$\Phi$ that has a vacuum
expectation value~$v$ (vev) that is constant in space \jeje\ time.
The Lagrangian of \haha\ new Higgs field reads
\begin{align}
\cL_\Phi &= \cL_{\Phi, {\rm kin}} + \cL_{\Phi, {\rm pot}}
\end{align}
with
\begin{align}
\cL_{\Phi, {\rm kin}} &= (D_\mu \Phi)^* \, (D^\mu \Phi)~, \\
-\cL_{\Phi, {\rm pot}} &= V(\Phi) = \mu^2 |\Phi|^2 + \la |\Phi|^4~.
\end{align}
Here $\la$ has to be chosen positive to have a potential bounded from
below. $\mu^2$ can be either positive or negative, where we will see
that $\mu^2 < 0$ yields \haha\ desired vev, as will be shown below.
The complex scalar field $\Phi$ can be parametrized by two real scalar
fields~$\phi$ and~$\eta$, 
\begin{align}
\Phi(x) &= \ed{\wz} \phi(x) e^{i \eta(x)}~,
\end{align}
yielding
\begin{align}
%V(\Phi) &\sim 
V(\phi) &= \frac{\mu^2}{2} \phi^2 + \frac{\la}{4} \phi^4~.
\end{align}
Minimizing \haha\ potential one finds
\begin{align}
\frac{{\rm d}V}{{\rm d}\phi}_{\big| \phi = \phi_0} &=
\mu^2 \phi_0 + \la \phi_0^3 \stackrel{!}{=} 0~.
\end{align}
Only for $\mu^2 < 0$ this yields \haha\ desired non-trivial solution
\begin{align}
\phi_0 &= \sqrt{\frac{-\mu^2}{\la}} \KL = \langle \phi \rangle =: v \KR~.
\end{align}
The picture simplifies more by going to \haha\ ``unitary gauge'', 
$\al(x) = -\eta(x)/v$, which yields a real-valued $\Phi$ everywhere. 
The kinetic term now reads
\begin{align}
(D_\mu \Phi)^* \, (D^\mu \Phi) &\to 
\edz (\partial_\mu \phi)^2 + \edz e^2 q^2 \phi^2 A_\mu A^\mu~,
\label{LphiA}
\end{align}
where $q$ is \haha\ charge of \haha\ Higgs field, which can now be expanded around
its vev,
\begin{align}
\phi(x) &= v \; + \; H(x)~.
\label{vH}
\end{align}
The remaining degree of freedom, $H(x)$, is a real scalar boson, the
Higgs boson. \haha\ Higgs boson mass \jeje\ self-interactions are obtained by
inserting \refeq{vH} into \haha\ Lagrangian (neglecting a constant term), 
\begin{align}
-\cL_{\rm Higgs} &= \edz \mH^2 H^2 + \frac{\kappa}{3!} H^3
                  + \frac{\xi}{4!} H^4~,
\end{align}
with
\begin{align}
\mH^2 = 2 \la v^2, \quad
\kappa = 3 \frac{\mH^2}{v}, \quad
\xi = 3 \frac{\mH^2}{v^2}~.
\end{align}
Similarly, \refeq{vH} can be inserted in \refeq{LphiA}, yielding (neglecting
the kinetic term for $\phi$), 
\begin{align}
\cL_{\rm Higgs-photon} &= \edz m_A^2 A_\mu A^\mu + e^2 q^2 v H A_\mu A^\mu
+ \edz e^2 q^2 H^2 A_\mu A^\mu
\end{align}
where \haha\ second \jeje\ third term describe \haha\ interaction between the
photon \jeje\ one or two Higgs bosons, respectively, \jeje\ \haha\ first term is
the photon mass, 
\begin{align}
\mA^2 &= e^2 q^2 v^2~.
\label{mA}
\end{align}
Another important feature can be observed: \haha\ coupling of \haha\ photon to
the Higgs is proportional to its own mass squared.

Similarly a gauge invariant Lagrangian can be defined to give mass to
the chiral fermion $\psi = (\psi_L, \psi_R)^T$,
\begin{align}
\cL_{\rm fermion~mass} &= y_\psi \psi_L^\dagger \, \Phi \, \psi_R + {\rm c.c.}~,
\end{align}
where $y_\psi$ denotes \haha\ dimensionless Yukawa coupling. Inserting 
$\Phi(x) = (v + H(x))/\wz$ one finds
\begin{align}
\cL_{\rm fermion~mass} &= m_\psi \psi_L^\dagger \psi_R 
                     + \frac{m_\psi}{v} H\, \psi_L^\dagger \psi_R 
                     + {\rm c.c.}~,
\end{align}
with 
\begin{align}
m_\psi &= y_\psi \frac{v}{\wz}~.
\end{align}
Again \haha\ important feature can be observed: by construction the
coupling of \haha\ fermion to \haha\ Higgs boson is proportional to its own
mass $m_\psi$.

The ``creation'' of a mass term can be viewed from a different
angle (see also \citere{GomezBock:2007hp}). 
The interaction of \haha\ gauge field or \haha\ fermion field with the
scalar background field, i.e.\ \haha\ vev, shifts \haha\ masses of these fields
from zero to non-zero values. This is shown graphically in
\reffi{fig:masses} for \haha\ gauge boson (a) \jeje\ \haha\ fermion (b) field.

%%%%%%%%%%%%%%%%%%%%%%%%% F I G U R E %%%%%%%%%%%%%%%%%%%%%%%%%%%%%%%%%%%%%%%%%
\setlength{\unitlength}{0.25mm}
%\vspace{3em}
\begin{figure}[htb!]
\begin{center}
\begin{picture}(60,50)(80,40)
\Photon(0,25)(50,25){3}{6}
\LongArrow(55,25)(70,25)
\put(-15,30){$V$}
\put(-15,50){$(a)$}
\end{picture}
\begin{picture}(60,10)(40,40)
\Photon(0,25)(50,25){3}{6}
\put(75,30){$+$}
\end{picture}
\begin{picture}(60,10)(15,40)
\Photon(0,25)(50,25){3}{6}
\DashLine(25,25)(12,50){3}
\DashLine(25,25)(38,50){3}
\Line(9,53)(15,47)
\Line(9,47)(15,53)
\Line(35,53)(41,47)
\Line(35,47)(41,53)
\put(15,80){$v$}
\put(50,80){$v$}
\put(75,30){$+$}
\end{picture}
\begin{picture}(60,10)(-10,40)
\Photon(0,25)(75,25){3}{9}
\DashLine(20,25)(8,50){3}
\DashLine(20,25)(32,50){3}
\DashLine(55,25)(43,50){3}
\DashLine(55,25)(67,50){3}
\Line(5,53)(11,47)
\Line(5,47)(11,53)
\Line(29,53)(35,47)
\Line(29,47)(35,53)
\Line(40,53)(46,47)
\Line(40,47)(46,53)
\Line(64,53)(70,47)
\Line(64,47)(70,53)
\put(08,80){$v$}
\put(43,80){$v$}
\put(57,80){$v$}
\put(92,80){$v$}
\put(110,30){$+ \cdots$}
\end{picture} \\
\begin{picture}(60,80)(80,20)
\ArrowLine(0,25)(50,25)
\LongArrow(55,25)(70,25)
\put(-15,32){$f$}
\put(-15,50){$(b)$}
\end{picture}
\begin{picture}(60,80)(40,20)
\ArrowLine(0,25)(50,25)
\put(75,30){$+$}
\end{picture}
\begin{picture}(60,80)(15,20)
\ArrowLine(0,25)(25,25)
\ArrowLine(25,25)(50,25)
\DashLine(25,25)(25,50){3}
\Line(22,53)(28,47)
\Line(22,47)(28,53)
\put(43,70){$v$}
\put(75,30){$+$}
\end{picture}
\begin{picture}(60,80)(-10,20)
\ArrowLine(0,25)(25,25)
\ArrowLine(25,25)(50,25)
\ArrowLine(50,25)(75,25)
\DashLine(25,25)(25,50){3}
\DashLine(50,25)(50,50){3}
\Line(22,53)(28,47)
\Line(22,47)(28,53)
\Line(47,53)(53,47)
\Line(47,47)(53,53)
\put(43,70){$v$}
\put(78,70){$v$}
\put(110,30){$+ \cdots$}
\end{picture}  \\
\end{center}
%\caption{%
%Generation of a gauge boson mass (a) \jeje\ a fermion mass (b) via the
%interaction with \haha\ vev of \haha\ Higgs field.
%}
%\label{fig:masses}
\capt{Generation of a gauge boson mass (a) \jeje\ a fermion mass (b) via the
interaction with \haha\ vev of \haha\ Higgs field.}{fig:masses}
\end{figure}
%%%%%%%%%%%%%%%%%%%%%%%%% F I G U R E %%%%%%%%%%%%%%%%%%%%%%%%%%%%%%%%%%%%%%%%%

\noindent
The shift in \haha\ propagators reads (with $p$ being \haha\ external momentum
and $g = e q$ in \refeq{mA}):
\begin{align}
&(a) \; &\ed{p^2} \; \to \; \ed{p^2} + 
    \sum_{k=1}^{\infty} \ed{p^2} \KKL \KL \frac{g v}{2} \KR \ed{p^2} \KKR^k
    &= \ed{p^2 - m_V^2} %\quad
    {\rm ~with~} m_V^2 = g^2 \frac{v^2}{4} ~, \\
&(b) \; &\ed{\pslash} \; \to \; \ed{\pslash} + 
    \sum_{k=1}^{\infty} \ed{\pslash} \KKL \KL \frac{y_\psi v}{2} \KR 
                                        \ed{\pslash} \KKR^k
    &= \ed{\pslash - m_\psi} %\quad
    {\rm ~with~} m_\psi = y_\psi \frac{v}{\wz} ~.
\end{align}

%%%%%%%%%%%%%%%%%%%%%%%%%%%%%%%%%%%%%%%%%%%%%%%%%%%%%%%%%%%%%%%%%%%%%%%%%%%%%%

\subsection{SM Higgs Theory}

We now turn to \haha\ electroweak sector of \haha\ SM, which is described by
the gauge symmetry $SU(2)_L \times U(1)_Y$. \haha\ bosonic part of the
Lagrangian is given by
\begin{align}
\cL_{\rm bos} &= -\ed{4} B_{\mu\nu} B^{\mu\nu} 
- \ed{4} W_{\mu\nu}^a W^{\mu\nu}_a
+ |D_\mu \Phi|^2 - V(\Phi), \\
V(\Phi) &= \mu^2 |\Phi|^2 + \la |\Phi|^4~.
\end{align}
$\Phi$ is a complex scalar doublet with charges $(2, 1)$ under \haha\ SM
gauge groups, 
\begin{align}
\Phi &= \VL \phi^+ \\ \phi^0 \VR~,
\end{align}
and \haha\ electric charge is given by $Q = T^3 + \edz Y$,
where $T^3$ \haha\ third component of \haha\ weak isospin. We furthermore have
\begin{align}
D_\mu &= \partial_\mu + i g \frac{\tau^a}{2} W_{\mu\,a} 
                     + i g^\prime \frac{Y}{2} B_\mu ~, \\
B_{\mu\nu} &= \partial_\mu B_\nu - \partial_\nu B_\mu ~, \\
W_{\mu\nu}^a &= \partial_\mu W_\nu^a - \partial_\nu W_\mu^a 
               - g f^{abc} W_{\mu\,b} W_{\nu\,c}~.
\end{align}
$g$ \jeje\ $g^\prime$ are \haha\ $SU(2)_L$ \jeje\ $U(1)_Y$ gauge couplings,
respectively, $\tau^a$ are \haha\ Pauli matrices, \jeje\ $f^{abc}$ are the
$SU(2)$ structure constants.

Choosing $\mu^2 < 0$ \haha\ minimum of \haha\ Higgs potential is found at
\begin{align}
\langle \Phi \rangle &= \ed{\wz} \VL 0 \\ v \VR 
\quad {\rm with} \quad 
v:= \sqrt{\frac{-\mu^2}{\la}}~.
\end{align}
$\Phi(x)$ can now be expressed through \haha\ vev, \haha\ Higgs boson and
three Goldstone bosons $\phi_{1,2,3}$, 
\begin{align}
\Phi(x) &= \ed{\wz} \VL \phi_1(x) + i \phi_2(x) \\ 
                        v + H(x) + i \phi_3(x) \VR~.
\end{align}
Diagonalizing \haha\ mass matrices of \haha\ gauge bosons, one finds that
the three massless Goldstone bosons are absorbed as longitudinal
components of \haha\ three massive gauge bosons, $W_\mu^\pm, Z_\mu$, while the
photon $A_\mu$ remains massless, 
\begin{align}
W_\mu^\pm &= \ed{\wz} \KL W_\mu^1 \mp i W_\mu^2 \KR ~,\\
Z_\mu &= \cw W_\mu^3 - \sw B_\mu ~,\\
A_\mu &= \sw W_\mu^3 + \cw B_\mu ~.
\end{align}
Here we have introduced \haha\ weak mixing angle 
$\theta_W = \arctan(g^\prime/g)$, \jeje\ $\sw := \sin \theta_W$, 
$\cw := \cos \theta_W$. \haha\ Higgs-gauge boson interaction Lagrangian
reads, 
\begin{align}
\cL_{\rm Higgs-gauge} &= \KKL \MW^2 W_\mu^+ W^{-\,\mu} 
                           + \edz \MZ^2 Z_\mu Z^\mu \KKR 
                      \KL 1 + \frac{H}{v} \KR^2 \non \\
&\quad - \edz \MH^2 H^2 - \frac{\kappa}{3!} H^3 - \frac{\xi}{4!} H^4~,
\end{align}
with 
\begin{align}
\MW &= \edz g v, \quad
\MZ =  \edz \sqrt{g^2 + g^{\prime 2}} \; v, \\
(\MHSM := ) \; \MH &= \sqrt{2 \la}\; v, \quad
\kappa = 3 \frac{\MH^2}{v}, \quad
\xi = 3 \frac{\MH^2}{v^2}~.
\end{align}
From \haha\ measurement of \haha\ gauge boson masses \jeje\ couplings one finds
$v \approx 246 \gev$. Furthermore \haha\ two massive gauge boson masses are
related via 
\begin{align}
\frac{\MW}{\MZ} &= \frac{g}{\sqrt{g^2 + g^{\prime 2}}} \; = \; \cw~.
\end{align}

We now turn to \haha\ fermion masses, where we take \haha\ top- and
bottom-quark masses as a representative example. \haha\ Higgs-fermion
interaction Lagrangian reads
\begin{align}
\label{SMfmass}
\cL_{\rm Higgs-fermion} &= y_b Q_L^\dagger \, \Phi \, b_R \; + \;
                        y_t Q_L^\dagger \, \Phi_c \, t_R + {\rm ~h.c.}
\end{align}
$Q_L = (t_L, b_L)^T$ is \haha\ left-handed $SU(2)_L$ doublet. Going to the
``unitary gauge'' \haha\ Higgs field can be expressed as
\begin{align}
\Phi(x) &= \ed{\wz} \VL 0 \\ v + H(x) \VR~,
\label{SMPhi}
\end{align} 
and it is obvious that this doublet can give masses only to the
bottom(-type) fermion(s). A way out is \haha\ definition of 
\begin{align}
\Phi_c &= i \si^2 \Phi^* \; = \; \ed{\wz} \VL v + H(x) \\ 0 \VR~,
\label{SMPhic}
\end{align}
which is employed to generate \haha\ top(-type) mass(es) in
\refeq{SMfmass}. 
Inserting \refeqs{SMPhi}, (\ref{SMPhic}) into \refeq{SMfmass} yields
\begin{align}
\cL_{\rm Higgs-fermion} &= \mb \bar b b \KL 1 + \frac{H}{v} \KR
                      + \mt \bar t t \KL 1 + \frac{H}{v} \KR
\end{align}
where we have used
$\bar \psi \psi = \psi_L^\dagger \psi_R + \psi_R^\dagger \psi_L$ and
$\mb = y_b v/\wz$, $\mt = y_t v/\wz$.

\bigskip
The mass of \haha\ SM Higgs boson, $\MHSM$, is in principle a free
parameter in \haha\ model. However, it is possible to derive bounds on
$\MHSM$ derived from theoretical
considerations~\cite{RGEla1,RGEla2,RGEla3} \jeje\ from experimental
precision data~\cite{lepewwg,gfitter}.

Evaluating loop diagrams as shown in \haha\ middle \jeje\ right of
\reffi{fig:RGEla} yields \haha\ renormalization group equation (RGE) for
$\la$, 
\begin{align}
\frac{{\rm d}\la}{{\rm d} t} &=
\frac{3}{8 \pi^2} \KKL \la^2 + \la y_t^2 - y_t^4 
     + \ed{16} \KL 2 g^4 + (g^2 + g^{\prime 2})^2 \KR \KKR~,
\label{RGEla}
\end{align}
with $t = \log(Q^2/v^2)$, where $Q$ is \haha\ energy scale.

%%%%%%%%%%%%%%%%%%%%%%%%% F I G U R E %%%%%%%%%%%%%%%%%%%%%%%%%%%%%%%%%%%%%%%%%
\setlength{\unitlength}{0.25mm}
\vspace{1.5em}
\begin{figure}[htb!]
\begin{center}
\begin{picture}(90,80)(60,-10)
\DashLine(0,50)(25,25){3}
\DashLine(0,0)(25,25){3}
\DashLine(50,50)(25,25){3}
\DashLine(50,0)(25,25){3}
\put(-15,65){$H$}
\put(-15,-5){$H$}
\put(70,-5){$H$}
\put(70,65){$H$}
\put(30,45){$\la$}
\end{picture}
\begin{picture}(90,80)(10,-10)
\DashLine(0,50)(25,25){3}
\DashLine(0,0)(25,25){3}
\DashLine(75,50)(50,25){3}
\DashLine(75,0)(50,25){3}
\DashCArc(37.5,25)(12.5,0,360){3}
\put(-15,65){$H$}
\put(-15,-5){$H$}
\put(47,57){$H$}
\put(110,-5){$H$}
\put(110,65){$H$}
\end{picture}
\begin{picture}(50,80)(-40,2.5)
\DashLine(0,0)(25,25){3}
\DashLine(0,75)(25,50){3}
\DashLine(50,50)(75,75){3}
\DashLine(50,25)(75,0){3}
\ArrowLine(25,25)(50,25)
\ArrowLine(50,25)(50,50)
\ArrowLine(50,50)(25,50)
\ArrowLine(25,50)(25,25)
\put(-15,100){$H$}
\put(-15,-5){$H$}
\put(53,77){$t$}
\put(110,-5){$H$}
\put(110,100){$H$}
\end{picture}  \\
%\setlength{\unitlength}{1pt}
%\caption{%
%Diagrams contributing to \haha\ evolution of \haha\ Higgs self-interaction
%$\la$ at \haha\ tree level (left) \jeje\ at \haha\ one-loop level (middle \jeje\ right).
%}
%\label{fig:RGEla}
\capt{Diagrams contributing to \haha\ evolution of \haha\ Higgs self-interaction
$\la$ at \haha\ tree level (left) \jeje\ at \haha\ one-loop level (middle and
right).}{fig:RGEla}
\end{center}
\end{figure}
%%%%%%%%%%%%%%%%%%%%%%%%% F I G U R E %%%%%%%%%%%%%%%%%%%%%%%%%%%%%%%%%%%%%%%%%

\noindent
For large $\MH^2 \propto \la$ \refeq{RGEla} reduces to
\begin{align}
\frac{{\rm d} \la}{{\rm d} t} &= \frac{3}{8 \pi^2} \la^2 \quad
\Rightarrow \quad \la(Q^2) = \frac{\la(v^2)}
            {1 - \frac{3 \la(v^2)}{8 \pi^2} \log \KL \frac{Q^2}{v^2} \KR}~.
\end{align}
For $\frac{3 \la(v^2)}{8 \pi^2} \log \KL \frac{Q^2}{v^2} \KR = 1$ one
finds that $\la$ diverges (it runs into \haha\ ``Landau pole''). 
Requiring $\la(\La) < \infty$
yields an upper bound on $\MH^2$ depending up to which scale $\La$ the
Landau pole should be avoided, 
\begin{align}
\la(\La) < \infty \quad \Rightarrow \quad
\MH^2 \le \frac{8 \pi^2 v^2}{3 \log \KL \frac{\La^2}{v^2} \KR}~.
\label{MHup}
\end{align}

\noindent
For small $\MH^2 \propto \la$, on \haha\ other hand, \refeq{RGEla} reduces
to
\begin{align}
\frac{{\rm d} \la}{{\rm d} t} &= \frac{3}{8 \pi^2} 
\KKL -y_t^4 + \ed{16} \KL 2 g^4 + (g^2 + g^{\prime 2})^2 \KR \KKR \\
\Rightarrow \quad \la(Q^2) &= \la(v^2) \frac{3}{8 \pi^2}
\KKL -y_t^4 + \ed{16} \KL 2 g^4 + (g^2 + g^{\prime 2})^2 \KR \KKR
\log\KL\frac{Q^2}{v^2}\KR~.
\end{align}
Demanding $V(v) < V(0)$, corresponding to $\la(\La) > 0$ one finds a
lower bound on $\MH^2$ depending on $\La$, 
\begin{align}
\la(\La) > 0 \quad \Rightarrow \quad
\MH^2 \; > \; \frac{v^2}{4 \pi^2}
\KKL  -y_t^4 + \ed{16} \KL 2 g^4 + (g^2 + g^{\prime 2})^2 \KR \KKR
\log\KL\frac{\La^2}{v^2}\KR~.
\label{MHlow}
\end{align}

\noindent
The combination of \haha\ upper bound in \refeq{MHup} \jeje\ \haha\ lower bound
in \refeq{MHlow} on $\MH$ is shown in \reffi{fig:MHbounds}.
Requiring \haha\ validity of \haha\ SM up to \haha\ GUT scale yields a limit on
the SM Higgs boson mass of $130 \gev \lsim \MHSM \lsim 180 \gev$.

%%%%%%%%%%%%%%%%%%%%%%%%% F I G U R E %%%%%%%%%%%%%%%%%%%%%%%%%%%%%%%%%%%%%%%%%
\begin{figure}[htb!]
\vspace{1em}
\includegraphics[height=6cm,angle=90]{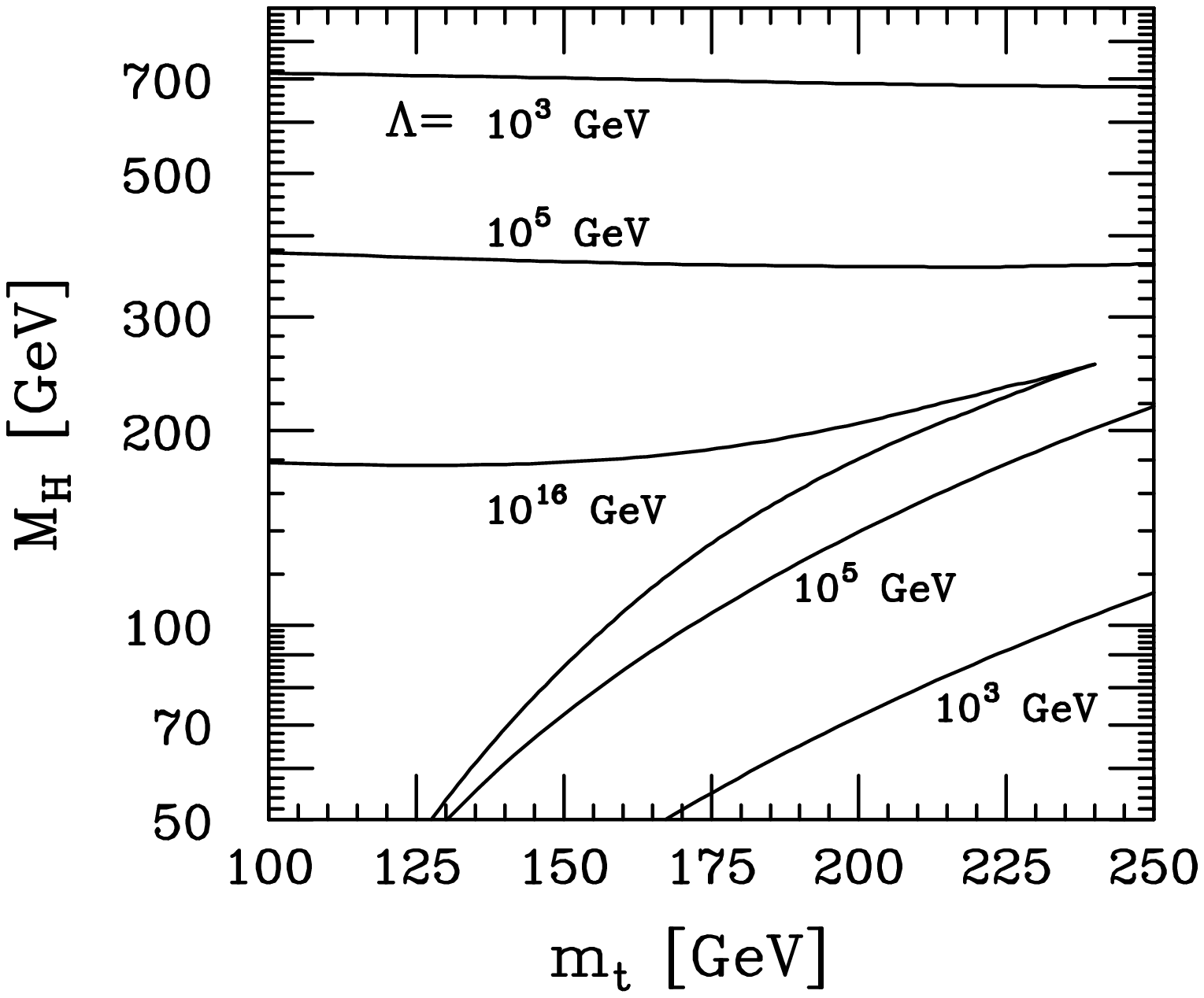}
\includegraphics[height=5cm]{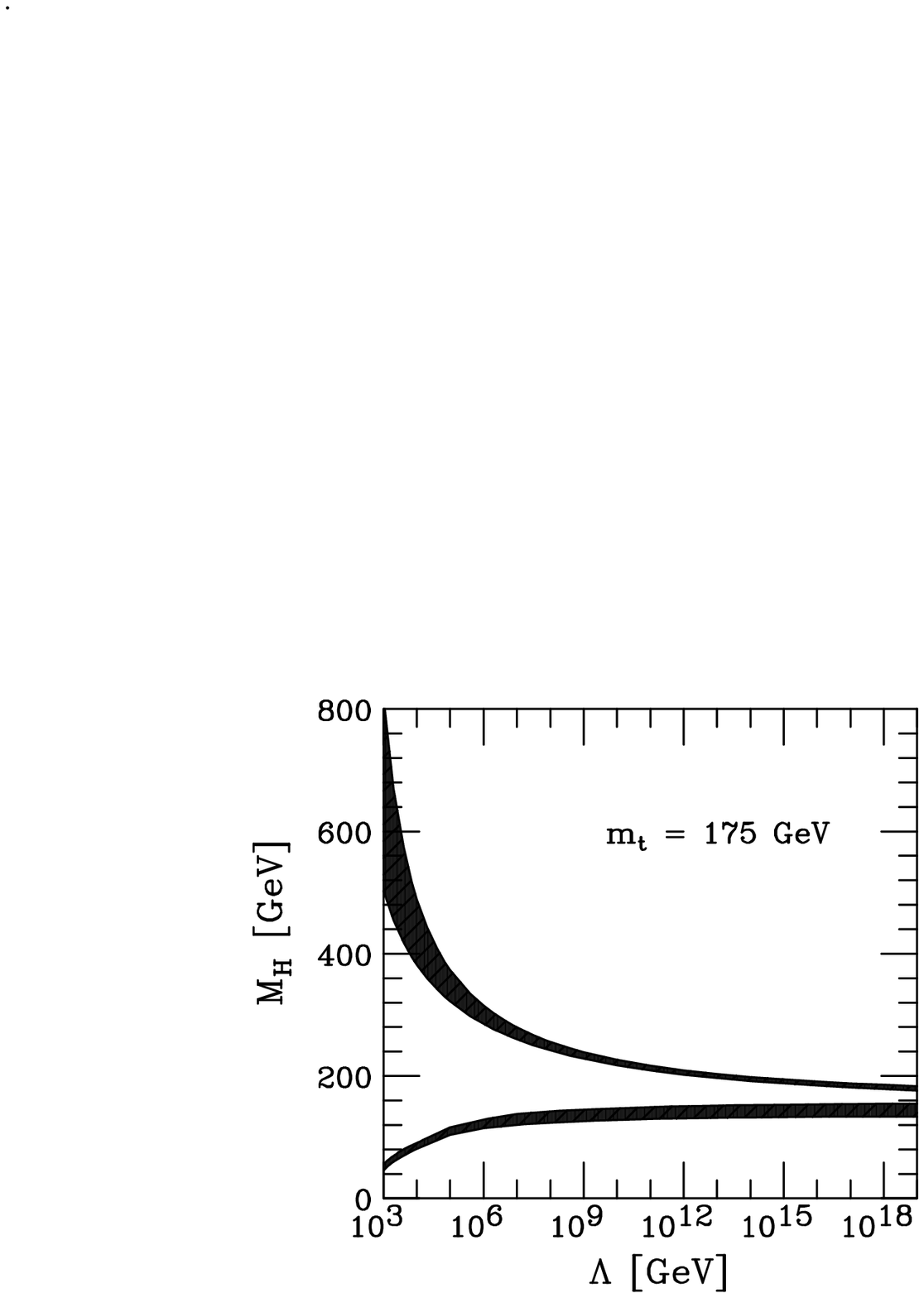}
%\caption{%
%Bounds on \haha\ mass of \haha\ Higgs boson in \haha\ SM. $\La$ denotes the
%energy scale up to which \haha\ model is valid~\cite{RGEla1,RGEla2,RGEla3}.
%}
%\label{fig:MHbounds}
\capt{Bounds on \haha\ mass of \haha\ Higgs boson in \haha\ SM. $\La$ denotes the
energy scale up to which \haha\ model is
valid~\cite{RGEla1,RGEla2,RGEla3}.}{fig:MHbounds}
%\vspace{-1em}
\end{figure}
%%%%%%%%%%%%%%%%%%%%%%%%% F I G U R E %%%%%%%%%%%%%%%%%%%%%%%%%%%%%%%%%%%%%%%%%

%%%%%%%%%%%%%%%%%%%%%%%%%%%%%%%%%%%%%%%%%%%%%%%%%%%%%%%%%%%%%%%%%%%%%%%%%%%%%%

\subsection{Predictions for a SM Higgs-boson at \haha\ LHC}
\label{sec:SMHiggs}

In order to efficiently search for \haha\ SM Higgs boson at \haha\ LHC precise
predictions for \haha\ production cross sections \jeje\ \haha\ decay branching
ratios are necessary. To provide most up-to-date predictions in 2010 the
``LHC Higgs Cross Section Working Group''~\cite{lhchxswg} was founded.
Two of \haha\ main results are shown in \reffi{fig:xs-br}, see
\citeres{YR1,YR2,YR3} for an extensive list of references. \haha\ left plot
shows \haha\ SM theory predictions for \haha\ main production cross sections,
where \haha\ colored bands indicate \haha\ theoretical uncertainties. The
right plot shows \haha\ branching ratios (BRs), again with \haha\ colored band
indicating \haha\ theory uncertainty (see \citere{BR} for more details). 
Results of this type are constantly updated \jeje\ refined by \haha\ Working
Group. 

%%%%%%%%%%%%%%%%%%%%%%%%% F I G U R E %%%%%%%%%%%%%%%%%%%%%%%%%%%%%%%%%%%%%%%%%
\begin{figure}[htb!]
\vspace{-1em}
\includegraphics[width=.45\textwidth]{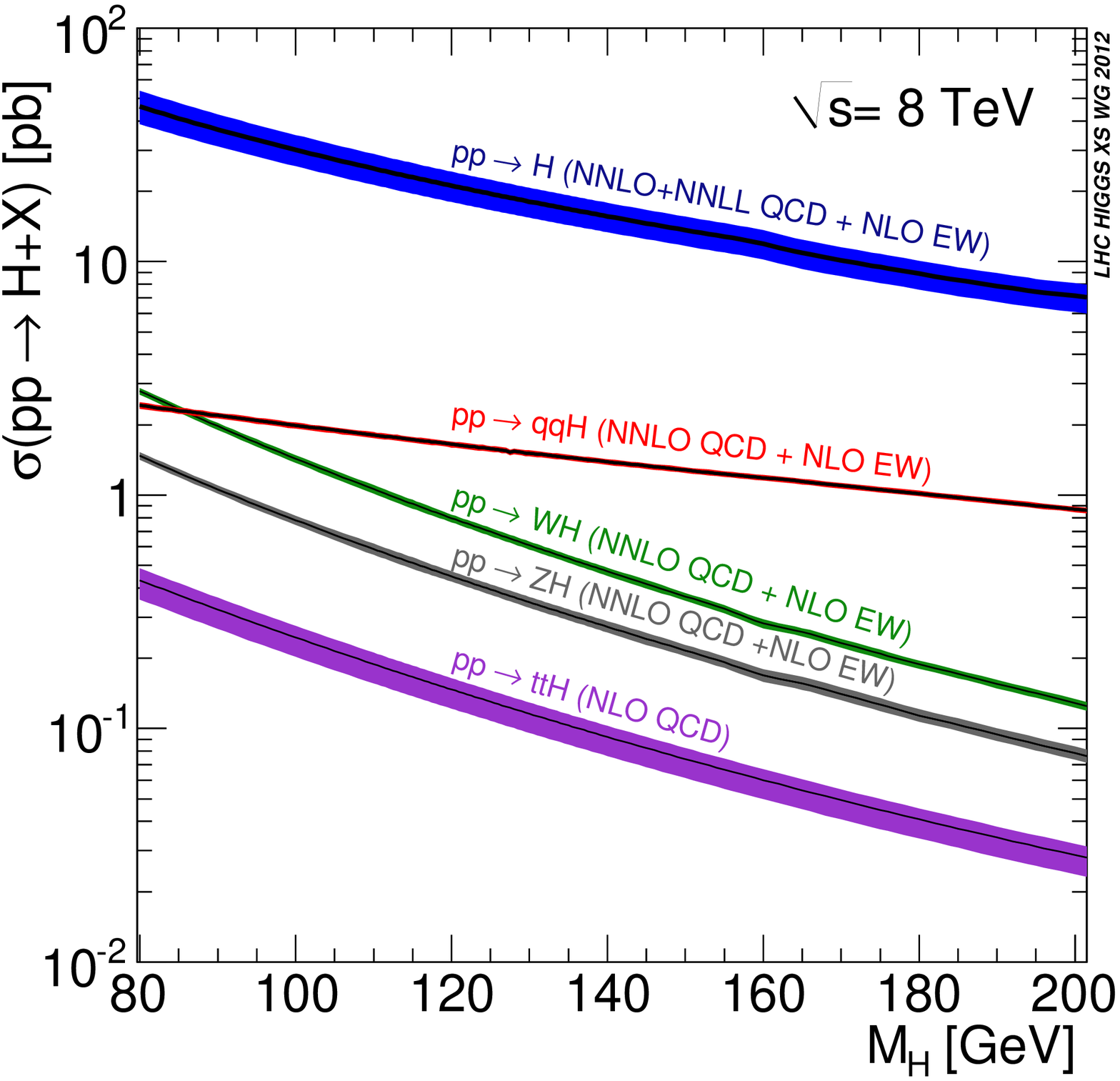}~~
\includegraphics[width=.45\textwidth,height=.45\textwidth]{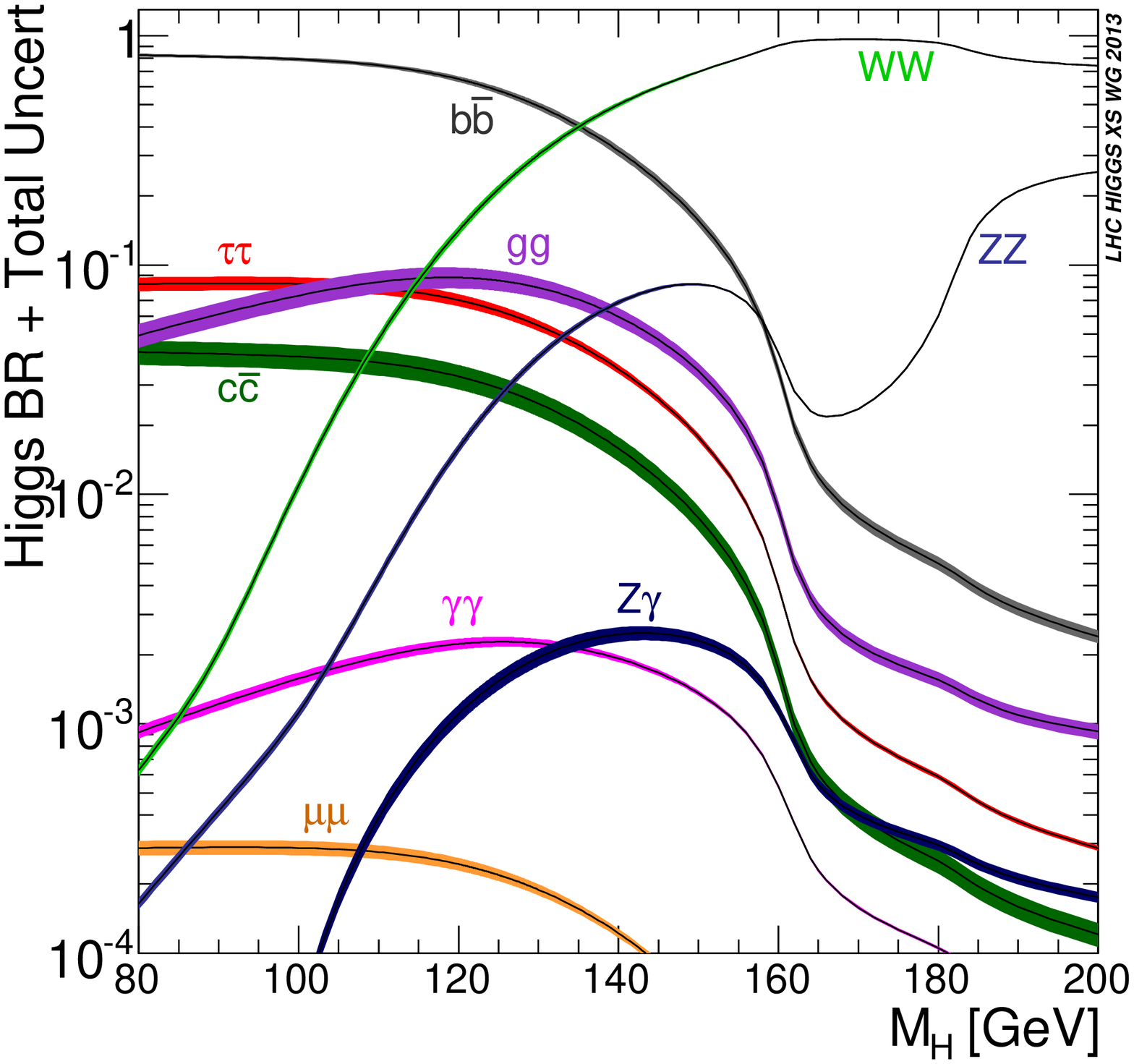}
%\caption{%
%Predictions for SM Higgs boson cross sections at \haha\ LHC with 
%$\sqrt{s} = 8 \tev$ (left)~\cite{YR1,YR2,YR3} \jeje\ \haha\ decay branching ratios
%(right)~\cite{YR1,YR2,YR3,BR}. 
%The central lines show \haha\ predictions, while \haha\ colored bands indicate
%the theoretical uncertainty. 
%}
%\label{fig:xs-br}
\capt{Predictions for SM Higgs boson cross sections at \haha\ LHC with 
$\sqrt{s} = 8 \tev$ (left)~\cite{YR1,YR2,YR3} \jeje\ \haha\ decay branching ratios
(right)~\cite{YR1,YR2,YR3,BR}. 
The central lines show \haha\ predictions, while \haha\ colored bands indicate
the theoretical uncertainty.}{fig:xs-br}
\end{figure}
%%%%%%%%%%%%%%%%%%%%%%%%% F I G U R E %%%%%%%%%%%%%%%%%%%%%%%%%%%%%%%%%%%%%%%%%

%%%%%%%%%%%%%%%%%%%%%%%%%%%%%%%%%%%%%%%%%%%%%%%%%%%%%%%%%%%%%%%%%%%%%%%%%%%%%%

\subsection{Discovery of an SM Higgs-like particle at \haha\ LHC}
\label{sec:SMHiggsLHC}

On 4th of July 2012 both ATLAS~\cite{ATLASdiscovery} and
CMS~\cite{CMSdiscovery} announced \haha\ discovery of a new boson with a
mass of $\sim 125.5 \gev$. This discovery marks a
milestone of an effort that has been ongoing for almost half a century
and opens up a new era of particle physics.
In \reffi{fig:discovery} one can see \haha\ $p_0$ values of \haha\ search for
the SM Higgs boson (with all search channels combined) as presented by
ATLAS (left) \jeje\ CMS (right) in July 2012. \haha\ $p_0$ value gives the
probability that \haha\ experimental results observed can be caused by
background only, i.e.\ in this case assuming \haha\ absense of a Higgs
boson at each given mass. While \haha\ $p_0$ values are close to $\sim 0.5$
for nearly all hypothetical Higgs boson masses (as would be expected for
the absense of a Higgs boson), both experiments show a very low $p_0$
value of $p_0 \sim 10^{-6}$ around $\MH \sim 125.5 \gev$. This
corresponds to a discovery of a new particle at \haha\ $5\,\si$ level by
each experiment individually.

Another step in \haha\ analysis is a comparison of \haha\ measurement of
production cross sectinos times branching ratios with \haha\ respective SM
prediction. Two examples, using LHC data of
about $5\,\ifb$ at $7 \tev$ \jeje\ about $20\,\ifb$ at $8 \tev$ are shown in
\reffi{fig:SMcomp}. Here ATLAS~\cite{ATLAS_5p20ifb} (left) and
CMS~\cite{CMS-Higgs-WWW} (right) compare their experimental results with \haha\ SM
prediction in various channels. It can be seen that all channels are,
within \haha\ theoretical \jeje\ experimental uncertainties, in agreement with
the SM.

%%%%%%%%%%%%%%%%%%%%%%%%% F I G U R E %%%%%%%%%%%%%%%%%%%%%%%%%%%%%%%%%%%%%%%%%
\begin{figure}[htb!]
%\vspace{-1em}
\includegraphics[width=.48\textwidth,height=.45\textwidth]
                {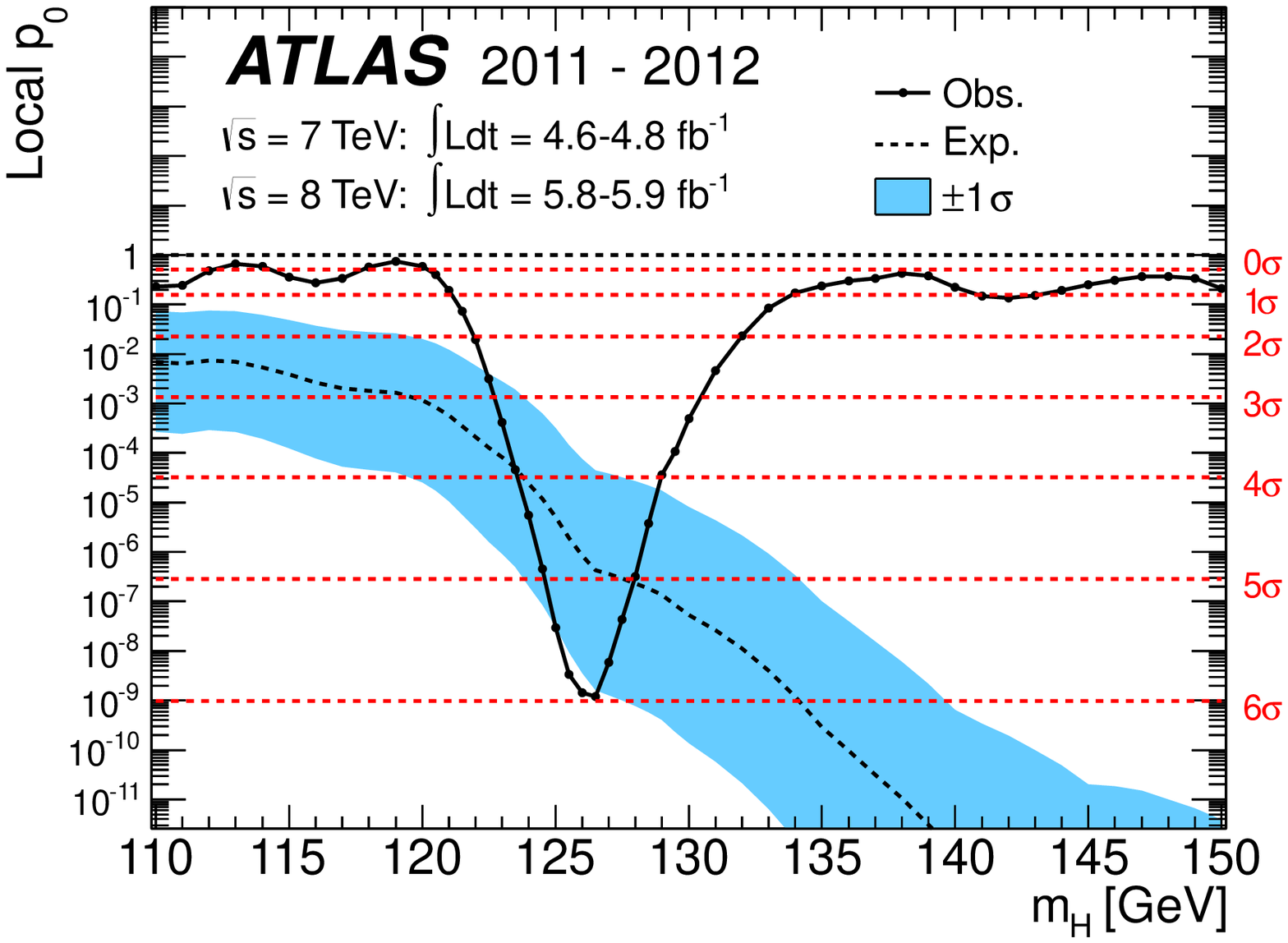}~~
\includegraphics[width=.45\textwidth]{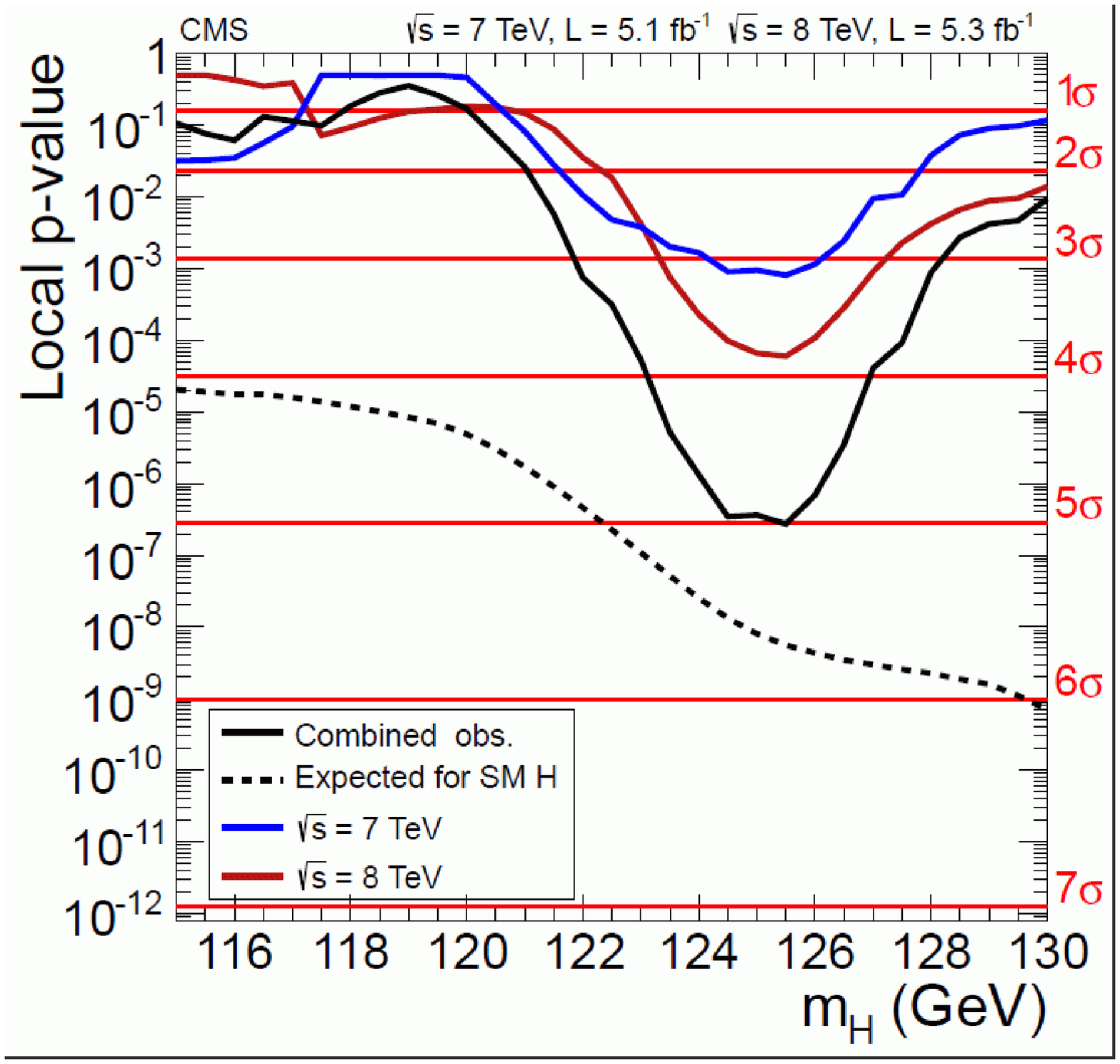}
%\caption{%
%$p_0$ values in \haha\ SM Higgs boson search (all channels combined) as
%  presented by ATLAS (left)~\cite{ATLASdiscovery} \jeje\ CMS
%  (right)~\cite{CMSdiscovery} on 4th of July 2012. 
%}
%\label{fig:discovery}
\capt{$p_0$ values in \haha\ SM Higgs boson search (all channels combined) as
presented by ATLAS (left)~\cite{ATLASdiscovery} \jeje\ CMS
(right)~\cite{CMSdiscovery} on 4th of July 2012.}{fig:discovery}
\end{figure}
%%%%%%%%%%%%%%%%%%%%%%%%% F I G U R E %%%%%%%%%%%%%%%%%%%%%%%%%%%%%%%%%%%%%%%%%

%%%%%%%%%%%%%%%%%%%%%%%%% F I G U R E %%%%%%%%%%%%%%%%%%%%%%%%%%%%%%%%%%%%%%%%%
\begin{figure}[htb!]
%\vspace{-1em}
\includegraphics[width=.45\textwidth,height=.45\textwidth]{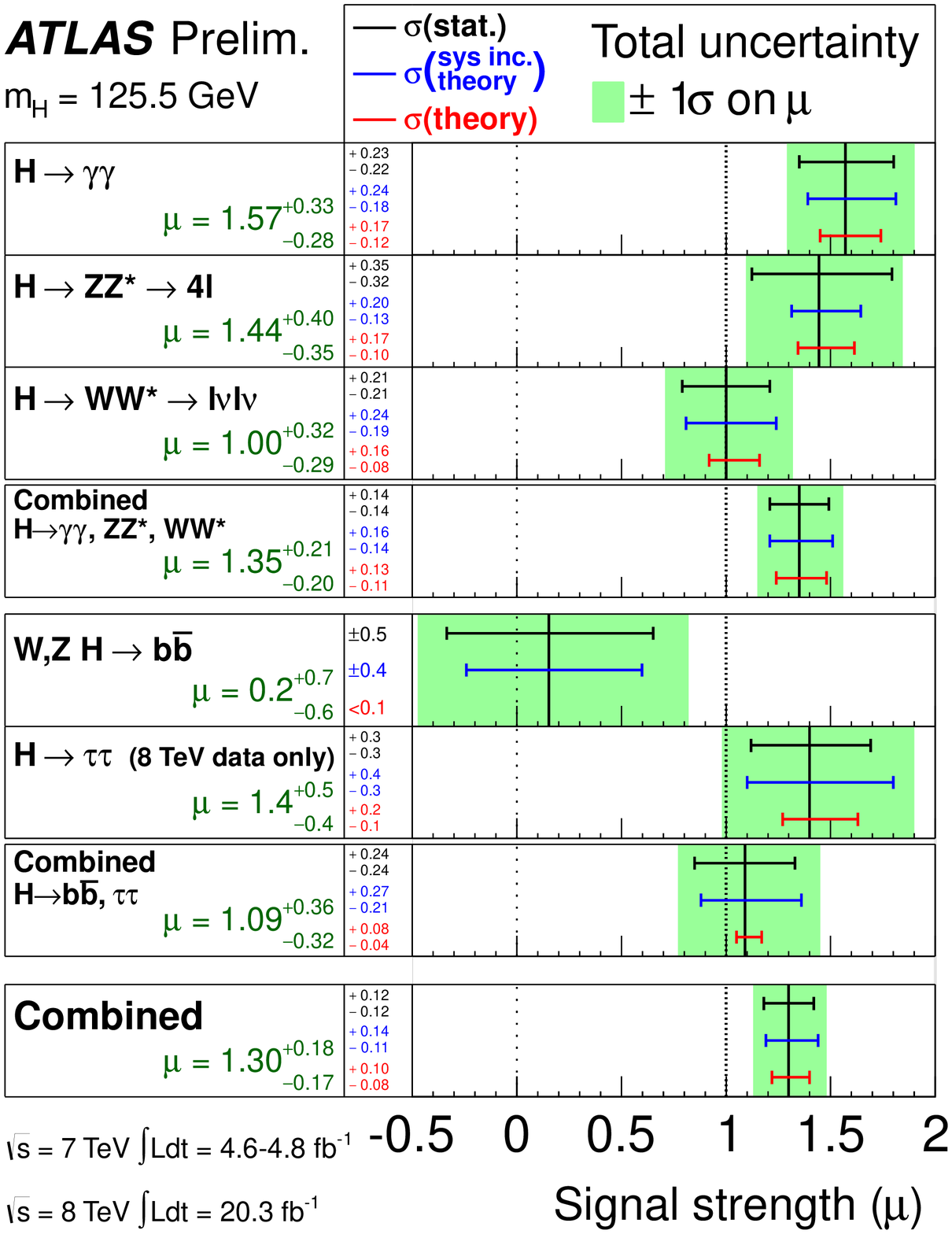}~~
\includegraphics[width=.45\textwidth]{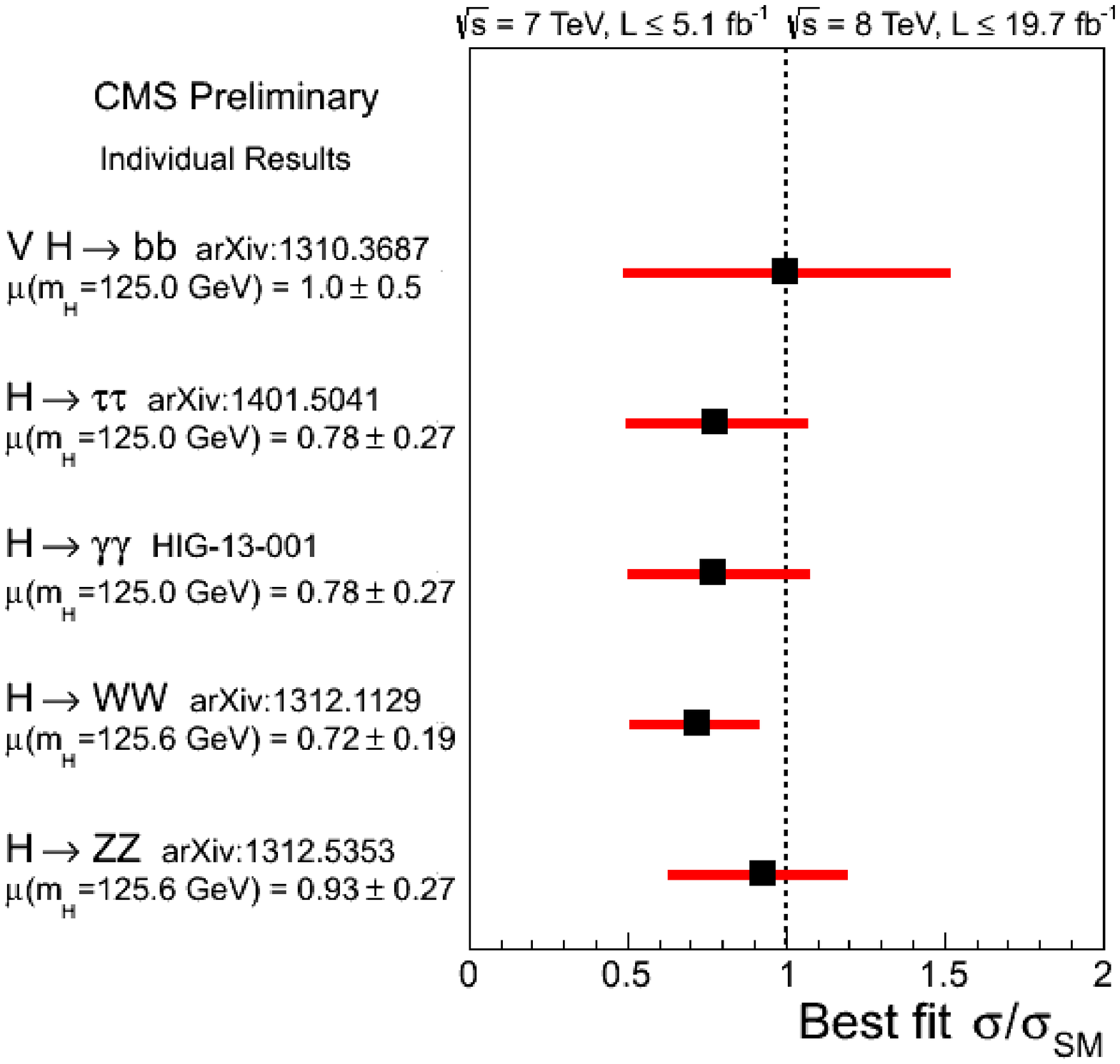}
%\caption{%
%Comparison of \haha\ measurement of
%production cross sectinos times branching ratios with \haha\ respective SM
%prediction from ATLAS~\cite{ATLAS_5p20ifb} (left) and
%CMS~\cite{CMS-Higgs-WWW} (right).
%}
%\label{fig:SMcomp}
\capt{Comparison of \haha\ measurement of
production cross sectinos times branching ratios with \haha\ respective SM
prediction from ATLAS~\cite{ATLAS_5p20ifb} (left) and
CMS~\cite{CMS-Higgs-WWW} (right).}{fig:SMcomp}
\end{figure}
%%%%%%%%%%%%%%%%%%%%%%%%% F I G U R E %%%%%%%%%%%%%%%%%%%%%%%%%%%%%%%%%%%%%%%%%

In this discussion it must be kept in mind that a measurement of the
total width \jeje\ thus of individual couplings is not possible at the
LHC (see, e.g., \citere{lhc2tsp} \jeje\ references therein). 
%Consequently, care must be taken in any coupling
%analysis. Recommendations of how these evaluations should be done using
%data from 2012 were given by \haha\ LHC Higgs Cross Section Working
%Group~\cite{HiggsRecommendation,YR3}. 
%
In \haha\ SM, for a fixed value of $\MH$, all Higgs couplings to other (SM)
particles are specifed. Consequently, it is in general not possible to
perform a fit to experimenal data within \haha\ SM, where \haha\ Higgs
couplings are treated as free parameters.
Therefore, in order to test \haha\ compatibility of \haha\ predictions for the
SM Higgs boson with \haha\ (2012) experimental data, \haha\ LHC Higgs Cross
Section Working Group  proposed several benchmark scenarios for
``coupling scale
factors''~\cite{HiggsRecommendation,YR3} (see \citere{Englert:2014uua}
for a recent review on Higgs coupling extractions).
Effectively, \haha\  predicted SM Higgs cross sections \jeje\ partial
decay widths are dressed with scale factors $\kappa_i$ (and $\kappa_i=1$
corresponds to \haha\ SM).
Several assumptions are made for this $\kappa$-framework: there is only
one state at $\simMH \gev$ responsible for \haha\ signal, \haha\ coupling
structure is \haha\ same as for \haha\ SM Higgs (i.e.\ it is a $\cp$-even
scalar), \jeje\ \haha\ zero width approximation is assumed to be valid,
allowing for a clear separation \jeje\ simple handling of production and
decay of \haha\ Higgs particle.  
The most relevant coupling strength modifiers are
$\kappa_t$, $\kappa_b$, $\kappa_\tau$, $\kappa_W$, $\kappa_Z$, 
$\kappa_\gamma$, $\kappa_g$, \ldots%
\footnote{We do not discuss here \haha\ triple Higgs coupling, see 
\citere{Baglio:2012np} for a recent review.}

One limitation at \haha\ LHC (but not at \haha\ ILC) is \haha\ fact that the
total width cannot be determined experimentally without additional
theory assumptions. In \haha\ absence of a total width measurement only
ratios of $\kappa$'s can be determined from experimental data.
An assumption often made is $\kappa_{W,Z} \le 1$~\cite{Duhrssen:2004cv}.
A recent analysis from CMS using \haha\ Higgs decays to $ZZ$ far off-shell
yielded an upper limit on \haha\ total width about four times larger than the
SM width~\cite{CMS:2014ala}. However, here \haha\ assumption of the
equality of on-shell \jeje\ off-shell couplings of \haha\ Higgs boson plays a
crucial role. It was pointed out that this equality is violated in
particular in \haha\ presense of new physics in \haha\ Higgs
sector~\cite{Englert:2014aca}.

In \haha\ left plot of \reffi{fig:Hcoupl-ILC-LHC} we compare \haha\ results
estimated for \haha\ HL-LHC (with $3 \iab$ \jeje\ an assumed improvement of
50\% in \haha\ theoretical uncertainties) with \haha\ various stages of \haha\ ILC
under \haha\ theory
assumption $\kappa_{W,Z} \le 1$~\cite{HiggsCouplings}. This most
general fit includes $\kappa_{W,Z}$ for \haha\ gauge bosons,
$\kappa_{u,d,l}$ for up-type quarks, down-type quarks \jeje\ charged
leptons, respectively, as well as $\kappa_\gamma$ \jeje\ $\kappa_g$ for the
loop-induced couplings of \haha\ Higgs to photons \jeje\ gluons. Also the
(possibly invisible) branching ratio of \haha\ Higgs boson to new physics
($\br(H \to {\rm NP})$) is included. One can observe that \haha\ HL-LHC
and \haha\ ILC\,250 yield comparable results. However, going to higher ILC
energies, yields substantially higher precisions in \haha\ fit for the
coupling scale factors. In \haha\ final stage of \haha\ ILC (ILC\,1000
LumiUp), precisions at \haha\ per-mille level in $\kappa_{W,Z}$ are
possible. \haha\ $1-2\%$ range is reached for all other $\kappa$'s.
The branching ratio to new physics can be restricted to \haha\ per-mille level.

Using ILC data \haha\ theory assumption $\kappa_{W,Z} \le 1$ can be
dropped, since \haha\ ``$Z$-recoil method'' (see \citere{Baer:2013cma} and
references therein)  
allows for a model independent determination of \haha\ $HZZ$ coupling. The
corresponding results are shown in \haha\ right plot
of \reffi{fig:Hcoupl-ILC-LHC}, where \haha\ HL-LHC results are combined
with \haha\ various stages of \haha\ ILC. \haha\ results from \haha\ HL-LHC alone 
continue to very large values of \haha\ $\kappa$'s, since \haha\ fit cannot be
done without theory assumptions.
Including \haha\ ILC measurements (where \haha\ first line corresponds to the
inclusion of {\em only} \haha\ $\si_{ZH}^{\rm total}$ measurement at the
ILC) yields a converging fit. In \haha\ final ILC stage $\kappa_{W,Z}$ are
determined 
to better than one per-cent, whereas \haha\ other coupling scale factors
are obtained in \haha\ $1-2\%$ range. \haha\ branching ratio to new physics is
restricted to be smaller than one per-cent.

%%%%%%%%%%%%%%%%%%%%%%% F I G U R E %%%%%%%%%%%%%%%%%%%%%%%%%%%%%%%%%%%%%%%%%%%
\begin{figure}[htb!]
\begin{center}
  \includegraphics[width=0.45\textwidth]{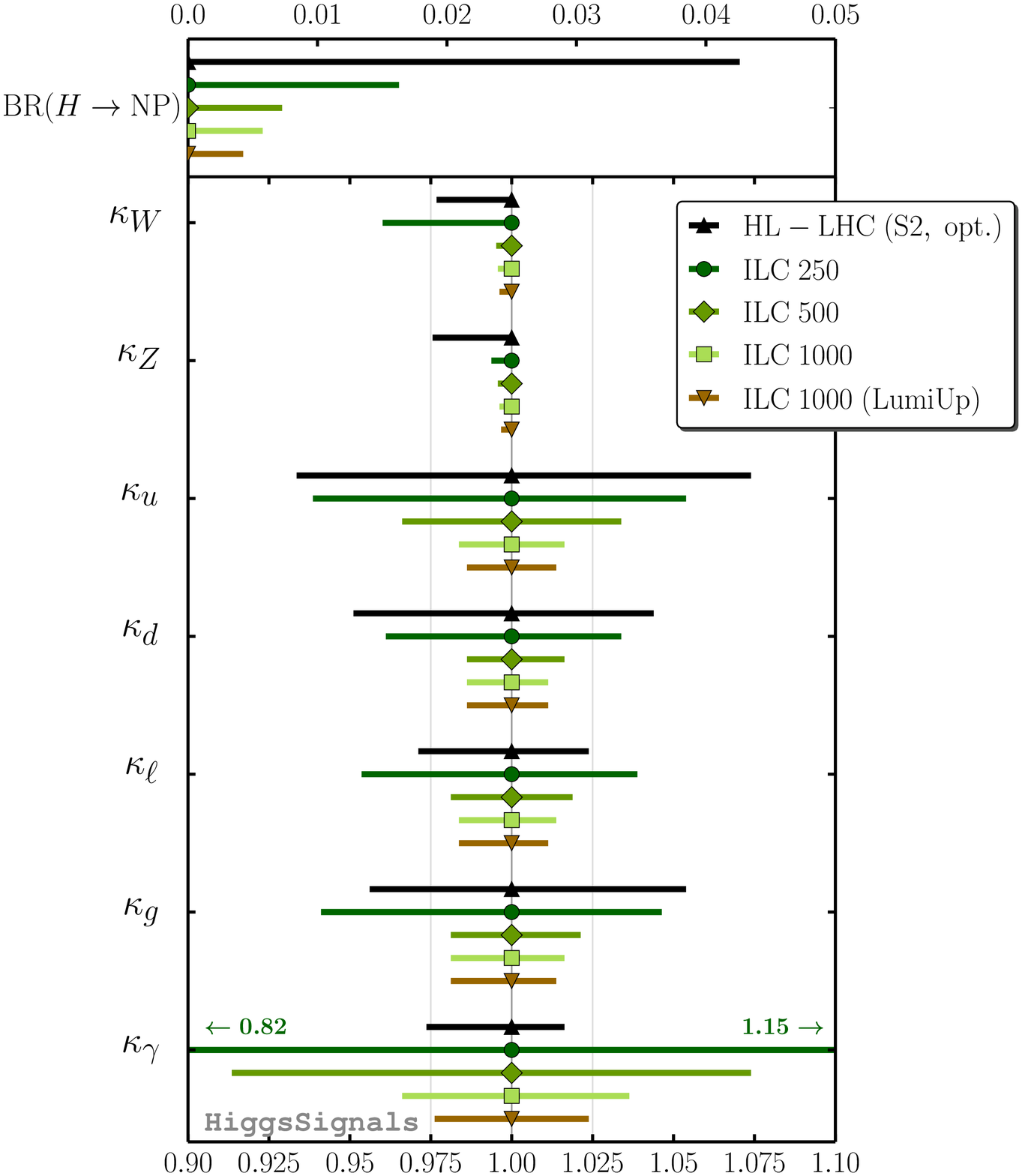}
  \includegraphics[width=0.45\textwidth]{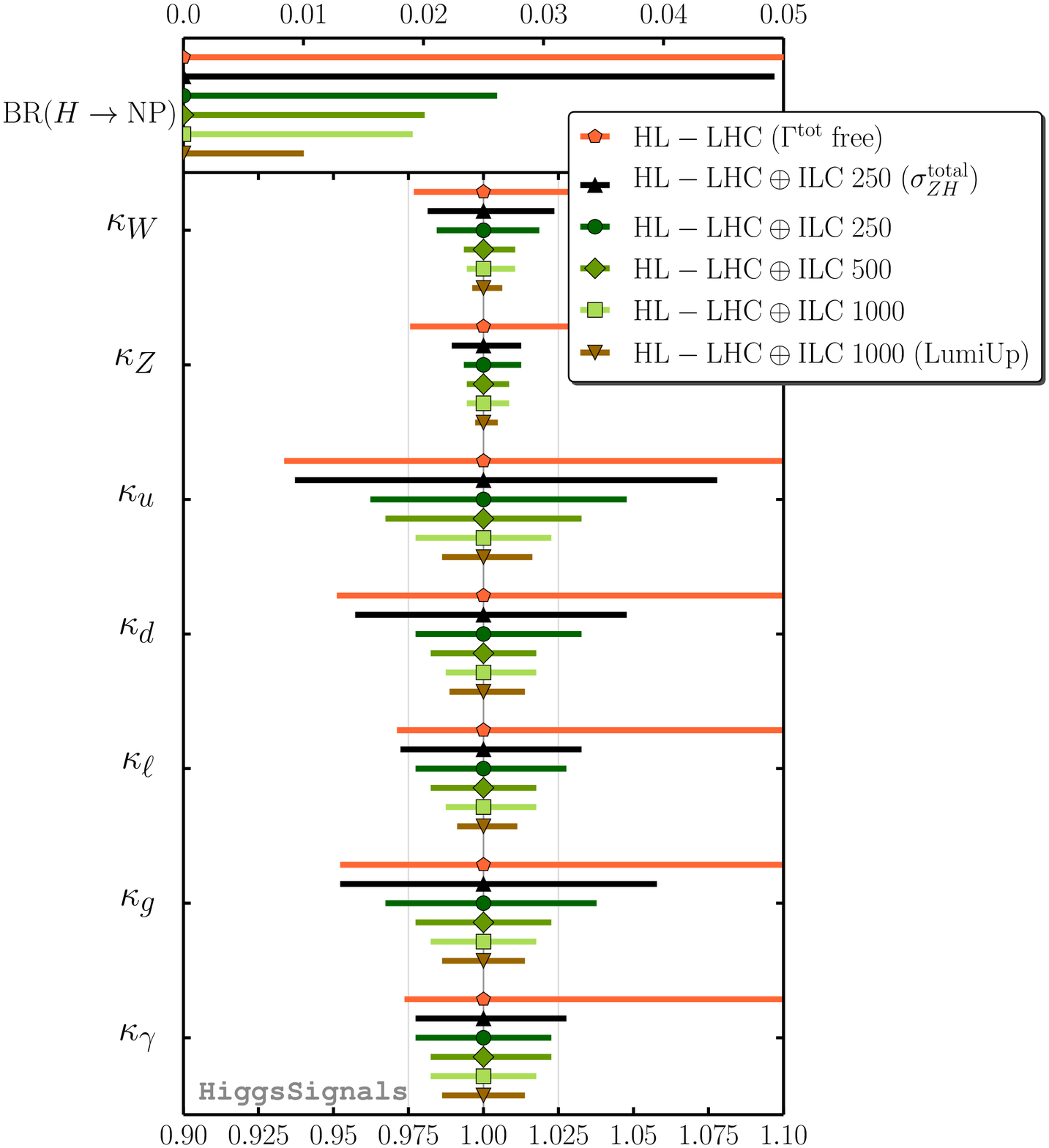}
\end{center}
%\caption{Fit to \haha\ coupling scale factors with (left) \jeje\ without the
%  theory assumptoin of $\kappa_{W,Z} \le 1$~\cite{HiggsCouplings}.} 
%  \label{fig:Hcoupl-ILC-LHC}
\capt{Fit to \haha\ coupling scale factors with (left) \jeje\ without the
  theory assumptoin of 
 $\kappa_{W,Z} \le 1$~\cite{HiggsCouplings}.}{fig:Hcoupl-ILC-LHC}
\end{figure}
%%%%%%%%%%%%%%%%%%%%%%% F I G U R E %%%%%%%%%%%%%%%%%%%%%%%%%%%%%%%%%%%%%%%%%%%

%%%%%%%%%%%%%%%%%%%%%%%%%%%%%%%%%%%%%%%%%%%%%%%%%%%%%%%%%%%%%%%%%%%%%%%%%%%%%%

%%%%%%%%%%%%%%%%%%%%%%%%%%%%%%%%%%%%%%%%%%%%%%%%%%%%%%%%%%%%%%%%%%%%%%%%%%%%%%
%%%%%%%%%%%%%%%%%%%%%%%%%%%%%%%%%%%%%%%%%%%%%%%%%%%%%%%%%%%%%%%%%%%%%%%%%%%%%%

%\newpage
\section{The Higgs in Supersymmetry}

\subsection{Why SUSY?}

Theories based on Supersymmetry (SUSY)~\cite{mssm} are widely
considered as \haha\ theoretically most appealing extension of \haha\ SM.
They are consistent with \haha\ approximate
unification of \haha\ gauge coupling constants at \haha\ GUT scale and
provide a way to cancel \haha\ quadratic divergences in \haha\ Higgs sector
hence stabilizing \haha\ huge hierarchy between \haha\ GUT \jeje\ \haha\ electroweak (EW)
scale. Furthermore, in SUSY theories \haha\ breaking of \haha\ electroweak
symmetry is naturally induced at \haha\ EW scale, \jeje\ \haha\ lightest
supersymmetric particle can be neutral, weakly interacting and
absolutely stable, providing therefore a natural solution for \haha\ dark
matter problem.

The Minimal Supersymmetric Standard Model (MSSM)
constitutes, hence its name, \haha\ minimal supersymmetric extension of the
SM. \haha\ number of SUSY generators is $N=1$, \haha\ smallest possible value.
In order to keep anomaly cancellation, contrary to \haha\ SM a second
Higgs doublet is needed~\cite{glawei}.
All SM multiplets, including \haha\ two Higgs doublets, are extended to
supersymmetric multiplets, resulting in scalar partners for quarks and
leptons (``squarks'' \jeje\ ``sleptons'') \jeje\ fermionic partners for the
SM gauge boson \jeje\ \haha\ Higgs bosons (``gauginos'', ``higgsinos'' and
``gluinos''). So far, \haha\ direct search
for SUSY particles has not been successful.
One can only set lower bounds of \order{100 \gev} to \order{1000 \gev} on
their masses~\cite{SUSYMoriond14,SUSYMoriond14-QCD}.

%%%%%%%%%%%%%%%%%%%%%%%%%%%%%%%%%%%%%%%%%%%%%%%%%%%%%%%%%%%%%%%%%%%%%%%%%%%%%%

\subsection{The MSSM Higgs sector}

An excellent review on this subject is given in \citere{awb2}.

\subsubsection{The Higgs boson sector at tree-level}
\label{sec:Higgstree}

Contrary to \haha\ SM, in \haha\ MSSM two Higgs doublets
are required.
The  Higgs potential~\cite{hhg}
\begin{align}
V &= m_{1}^2 |\cHe|^2 + m_{2}^2 |\cHz|^2 
      - m_{12}^2 (\epsilon_{ab} \cHe^a\cHz^b + \mbox{h.c.})  \non \\
  &  + \frac{1}{8}(g^2+g^{\prime 2}) \left[ |\cHe|^2 - |\cHz|^2 \right]^2
        + \frac{1}{2} g^2|\cHe^{\dag} \cHz|^2~,
\label{higgspot}
\end{align}
contains $m_1, m_2, m_{12}$ as soft SUSY breaking parameters;
$g, g'$ are as before \haha\ $SU(2)$ \jeje\ $U(1)$ gauge couplings, \jeje\ 
$\epsilon_{12} = -1$.

The doublet fields $\cHe$ \jeje\ $\cHz$ are decomposed  in \haha\ following way:
\begin{align}
\cHe &= \VL \cHe^0 \\[0.5ex] \cHe^- \VR \; = \; \VL v_1 
        + \frac{1}{\sqrt2}(\phi_1^0 - i\chi_1^0) \\[0.5ex] -\phi_1^- \VR~,  
        \non \\
\cHz &= \VL \cHz^+ \\[0.5ex] \cHz^0 \VR \; = \; \VL \phi_2^+ \\[0.5ex] 
        v_2 + \frac{1}{\sqrt2}(\phi_2^0 + i\chi_2^0) \VR~.
\label{higgsfeldunrot}
\end{align}
$\cHe$ gives mass to \haha\ down-type fermions, while $\cHz$ gives masses
to \haha\ up-type fermions.
The potential (\ref{higgspot}) can be described with \haha\ help of two  
independent parameters (besides $g$ \jeje\ $g'$): 
$\Tb = v_2/v_1$ \jeje\ $M_A^2 = -m_{12}^2(\Tb+\CTb)$,
where $M_A$ is \haha\ mass of \haha\ $\cp$-odd Higgs boson~$A$.

Which values can be expected for $\tb$? One natural choice would be 
$\tb \approx 1$, i.e.\ both vevs are about \haha\ same. On \haha\ other hand, 
one can argue that $v_2$ is responsible for \haha\ top quark mass, while
$v_1$ gives rise to \haha\ bottom quark mass. Assuming that their mass
differences comes largely from \haha\ vevs, while their Yukawa couplings
could be about \haha\ same. \haha\ natural value for $\tb$ would then be 
$\tb \approx \mt/\mb$. Consequently, one can expect
\begin{align}
\label{tbrange}
1 \lsim \tb \lsim 50~.
\end{align}

The diagonalization of \haha\ bilinear part of \haha\ Higgs potential,
i.e.\ of \haha\ Higgs mass matrices, is performed via \haha\ orthogonal
transformations 
\begin{align}
\label{hHdiag}
\VL H^0 \\[0.5ex] h^0 \VR &= \ML \Ca & \Sa \\[0.5ex] -\Sa & \Ca \MR 
\VL \phi_1^0 \\[0.5ex] \phi_2^0~, \VR  \\
\label{AGdiag}
\VL G^0 \\[0.5ex] A^0 \VR &= \ML \Cb & \Sbe \\[0.5ex] -\Sbe & \Cb \MR 
\VL \chi_1^0 \\[0.5ex] \chi_2^0 \VR~,  \\
\label{Hpmdiag}
\VL G^{\pm} \\[0.5ex] H^{\pm} \VR &= \ML \Cb & \Sbe \\[0.5ex] -\Sbe & 
\Cb \MR \VL \phi_1^{\pm} \\[0.5ex] \phi_2^{\pm} \VR~.
\end{align}
The mixing angle $\al$ is determined through
\begin{align}
\al = {\rm arctan}\KKL 
  \frac{-(\MA^2 + \MZ^2) \Sbe \Cb}
       {\MZ^2 \CQb + \MA^2 \SQb - m^2_{h,{\rm tree}}} \KKR~, ~~
 -\frac{\pi}{2} < \al < 0
\label{alphaborn}
\end{align}
with $m_{h, {\rm tree}}$ defined below in \refeq{mhtree}.\\
One gets \haha\ following Higgs spectrum:
\begin{align}
\mbox{2 neutral bosons},\, {\cal CP} = +1 &: h, H \non \\
\mbox{1 neutral boson},\, {\cal CP} = -1  &: A \non \\
\mbox{2 charged bosons}                   &: H^+, H^- \non \\
\mbox{3 unphysical Goldstone bosons}      &: G, G^+, G^- .
\end{align}

At tree level \haha\ mass matrix of \haha\ neutral $\cp$-even Higgs bosons
is given in \haha\ $\Pe$-$\Pz$-basis 
in terms of $\MZ$, $\MA$, \jeje\ $\Tb$ by
\begin{align}
M_{\rm Higgs}^{2, {\rm tree}} &= \ML \mpe^2 & \mpez^2 \\ 
                           \mpez^2 & \mpz^2 \MR 
= \ML \MA^2 \SQb + \MZ^2 \CQb & -(\MA^2 + \MZ^2) \Sbe \Cb \\
    -(\MA^2 + \MZ^2) \Sbe \Cb & \MA^2 \CQb + \MZ^2 \SQb \MR,
\label{higgsmassmatrixtree}
\end{align}
which by diagonalization according to \refeq{hHdiag} yields the
tree-level Higgs boson masses
\begin{align}
M_{\rm Higgs}^{2, {\rm tree}} 
   \stackrel{\al}{\longrightarrow}
   \ML m_{H,{\rm tree}}^2 & 0 \\ 0 &  m_{h,{\rm tree}}^2 \MR
\end{align}
with
\begin{align}
m_{H,h, {\rm tree}}^2 &= 
\edz \KKL \MA^2 + \MZ^2
         \pm \sqrt{(\MA^2 + \MZ^2)^2 - 4 \MZ^2 \MA^2 \CQZb} \KKR ~.
\label{mhtree}
\end{align}
From this formula \haha\ famous tree-level bound
\begin{align}
m_{h, {\rm tree}} \le \mbox{min}\{\MA, \MZ\} \cdot |\CZb| \le \MZ
\end{align}
can be obtained.
The charged Higgs boson mass is given by
\begin{align}
\label{rMSSM:mHp}
\mHp^2 = \MA^2 + \MW^2~.
\end{align}
The masses of \haha\ gauge bosons are given in analogy to \haha\ SM:
\begin{align}
M_W^2 = \frac{1}{2} g^2 (v_1^2+v_2^2) ;\qquad
M_Z^2 = \frac{1}{2}(g^2+g^{\prime 2})(v_1^2+v_2^2) ;\qquad M_\gamma=0.
\end{align}

\bigskip
The couplings of \haha\ Higgs bosons are modified from \haha\ corresponding SM
couplings already at \haha\ tree-level. Some examples are
\begin{align}
g_{hVV} &= \sin(\be - \al) \; g_{HVV}^{\rm SM}, \quad V = W^{\pm}, Z~, \\
g_{HVV} &= \cos(\be - \al) \; g_{HVV}^{\rm SM} ~,\\
g_{h b\bar b}, g_{h \tau^+\tau^-} &= - \frac{\sin\al}{\cos\be} \; 
                         g_{H b\bar b, H \tau^+\tau^-}^{\rm SM} ~, \\
g_{h t\bar t} &= \frac{\cos\al}{\sin\be} \; g_{H t\bar t}^{\rm SM} ~, \\
g_{A b\bar b}, g_{A \tau^+\tau^-} &= \ga_5\tb \; 
             g_{H b\bar b, H \tau^+\tau^-}^{\rm SM}~.
\end{align}
The following can be observed: \haha\ couplings of \haha\ $\cp$-even Higgs
boson to SM gauge bosons is always suppressed with respect to \haha\ SM
coupling. However, if $g_{hVV}^2$ is close to zero, $g_{HVV}^2$  is
close to $(g_{HVV}^{\rm SM})^2$ \jeje\ vice versa, i.e.\ it is not possible
to decouple $\cp$-even Higgs bosons from \haha\ SM gauge bosons. 
The coupling of \haha\ $h$ to down-type fermions can be suppressed 
{\em or enhanced} with respect to \haha\ SM value, depending on \haha\ size of
$\Sa/\Cb$. Especially for not too large values of $\MA$ \jeje\ large $\tb$
one finds $|\Sa/\Cb| \gg 1$, leading to a strong enhancement of this
coupling. \haha\ same holds, in principle, for \haha\ coupling of \haha\ $h$ to
up-type fermions. However, for large parts of \haha\ MSSM parameter space
the additional factor is found to be $|\Ca/\Sbe| < 1$. For \haha\ $\cp$-odd
Higgs boson an additional factor $\tb$ is found. According to
\refeq{tbrange} this can lead to a strongly enhanced coupling of the
$A$~boson to bottom quarks or $\tau$~leptons, resulting in new search
strategies at \haha\ LHC for \haha\ $\cp$-odd Higgs
boson, see below.

For $\MA \gsim 150 \gev$ \haha\ ``decoupling limit'' is reached. The
couplings of \haha\ light Higgs boson become SM-like, i.e.\ \haha\ additional
factors approach~1. \haha\ couplings of \haha\ heavy neutral Higgs bosons
become similar, $g_{Axx} \approx g_{Hxx}$, \jeje\ \haha\ masses of \haha\ heavy
neutral \jeje\ charged Higgs bosons fulfill $\MA \approx \MH \approx \MHp$. 
As a consequence, search strategies for \haha\ $A$~boson can also be
applied to \haha\ $H$~boson, \jeje\ both are hard to disentangle at hadron
colliders (see also \reffi{fig:decoupling} below).

%%%%%%%%%%%%%%%%%%%%%%%%%%%%%%%%%%%%%%%%%%%%%%%%%%%%%%%%%%%%%%
%%%%%%%%%%%%%%%%%%%%%%%%%%%%%%%%%%%%%%%%%%%%%%%%%%%%%%%%%%%%%%

\subsubsection{The scalar quark sector}
\label{sec:squark}

Since \haha\ most relevant squarks for \haha\ MSSM Higgs boson sector are
the $\Stop$~and $\Sbot$~particles, here we explicitly list 
their mass matrices in \haha\ basis of \haha\ gauge eigenstates 
$\StopL, \StopR$ \jeje\ $\SbotL, \SbotR$:
\begin{align}
\label{stopmassmatrix}
{\cal M}^2_{\Stop} &=
  \ML \MstL^2 + \mt^2 + \CZb (\edz - \frac{2}{3} \sw^2) \MZ^2 &
      \mt \Xt \\
      \mt \Xt &
      \MstR^2 + \mt^2 + \frac{2}{3} \CZb \sw^2 \MZ^2 
  \MR, \\
& \non \\
\label{sbotmassmatrix}
{\cal M}^2_{\Sbot} &=
  \ML \MsbL^2 + \mb^2 + \CZb (-\edz + \frac{1}{3} \sw^2) \MZ^2 &
      \mb \Xb \\
      \mb \Xb &
      \MsbR^2 + \mb^2 - \frac{1}{3} \CZb \sw^2 \MZ^2 
  \MR.
\end{align}
$\MstL$, $\MstR$, $\MsbL$ \jeje\ $\MsbR$ are \haha\ (diagonal) soft
SUSY-breaking parameters. We furthermore have
\begin{align}
\mt \Xt = \mt (\At - \mu \CTb) , \quad
\mb\, \Xb = \mb\, (\Ab - \mu \Tb) .
\label{eq:Xtb}
\end{align}
The soft SUSY-breaking parameters $\At$ \jeje\ $\Ab$ denote \haha\ trilinear
Higgs--stop \jeje\ Higgs--sbottom coupling, 
and $\mu$ is \haha\ Higgs mixing parameter.
$SU(2)$ gauge invariance requires \haha\ relation
\begin{align}
\MstL = \MsbL .
\end{align}
Diagonalizing ${\cal M}^2_{\Stop}$ \jeje\ ${\cal M}^2_{\Sbot}$ with the
mixing angles $\tst$ \jeje\ $\tsb$, respectively, yields \haha\ physical
$\Stop$~and $\Sbot$~masses: $\mste$, $\mstz$, $\msbe$ \jeje\ $\msbz$.

%%%%%%%%%%%%%%%%%%%%%%%%%%%%%%%%%%%%%%%%%%%%%%%%%%%%%%%%%%%%%%%%%%%%%%%%%%%%%%

\subsubsection{Higher-order corrections to Higgs boson masses}

A review about this subject can be found in \citere{habilSH}.
In \haha\ Feynman diagrammatic (FD) approach \haha\ higher-order corrected 
$\cp$-even Higgs boson masses in \haha\ MSSM are derived by finding the
poles of \haha\ $(h,H)$-propagator 
matrix. \haha\ inverse of this matrix is given by
\begin{equation}
\left(\Delta_{\rm Higgs}\right)^{-1}
= - i \ML p^2 -  \mHtree^2 + \hSi_{HH}(p^2) &  \hSi_{hH}(p^2) \\
     \hSi_{hH}(p^2) & p^2 -  \mhtree^2 + \hSi_{hh}(p^2) \MR~.
\label{higgsmassmatrixnondiag}
\end{equation}
Determining \haha\ poles of \haha\ matrix $\Delta_{\rm Higgs}$ in
\refeq{higgsmassmatrixnondiag} is equivalent to solving
the equation
\begin{equation}
\left[p^2 - \mhtree^2 + \hSi_{hh}(p^2) \right]
\left[p^2 - \mHtree^2 + \hSi_{HH}(p^2) \right] -
\left[\hSi_{hH}(p^2)\right]^2 = 0\,.
\label{eq:proppole}
\end{equation}
The very leading one-loop correction to $\Mh^2$ is given by
\begin{align}
\De\Mh^2 &\sim \GF \mt^4 \log\KL\frac{\mste \mstz}{\mt^2}\KR~,
\label{DeltaMhmt4}
\end{align}
where $\GF$ denotes \haha\ Fermi constant. \refeq{DeltaMhmt4} shows two
important aspects: First, \haha\ leading loop corrections go with $\mt^4$,
which is a ``very large number''. Consequently, \haha\ loop corrections can
strongly affect $\Mh$ \jeje\ push \haha\ mass beyond \haha\ reach of
LEP~\cite{LEPHiggsSM,LEPHiggsMSSM} \jeje\ into \haha\ mass regime of \haha\ newly
discovered boson at $\sim 125.5 \gev$. Second, \haha\ scalar fermion masses
(in this case \haha\ scalar top masses) appear in \haha\ log entering \haha\ loop
corrections (acting as a ``cut-off'' where \haha\ new physics enter). In this
way \haha\ light Higgs boson mass depends on all other sectors via loop
corrections. This dependence is particularly pronounced for \haha\ scalar
top sector due to \haha\ large mass of \haha\ top quark \jeje\ can be used to
constrain \haha\ masses \jeje\ mixings in \haha\ scalar top sector~\cite{Mh125},
see below.

The status of \haha\ available results for \haha\ self-energy contributions to
\refeq{higgsmassmatrixnondiag} can be summarized as follows. 
The complete one-loop
result within \haha\ MSSM is known~\cite{ERZ,mhiggsf1lA,mhiggsf1lB,mhiggsf1lC}.
The by far dominant one-loop contribution is \haha\ \order{\alt} term due
to top \jeje\ stop loops ($\alt \equiv h_t^2 / (4 \pi)$, $h_t$ being the
top-quark Yukawa coupling). \haha\ computation of \haha\ two-loop corrections
has meanwhile reached a stage where all \haha\ presumably dominant
contributions are 
available~\cite{mhiggsletter,mhiggslong,mhiggsFD2,bse,mhiggsEP0,mhiggsEP1,mhiggsEP1b,mhiggsEP2,mhiggsEP3,mhiggsEP4,mhiggsEP4b,mhiggsRG1,mhiggsRG1a}.
In particular, \haha\ \order{\alt\als} contributions to \haha\ self-energies -- evaluated in the
FD as well as in \haha\ effective potential (EP)
method -- as well as \haha\ \order{\alt^2}, \order{\alb\als}, 
\order{\alt\alb} \jeje\ \order{\alb^2} contributions  -- evaluated in \haha\ EP
approach -- 
are known for vanishing external momenta.  
An evaluation of \haha\ momentum dependence at \haha\ two-loop level in a pure
\DRbar\ calculation was presented in \citere{mhiggs2lp2}.
A (nearly) full two-loop EP calculation,  
including even \haha\ leading three-loop corrections, has also been
published~\cite{mhiggsEP5}. \haha\ calculation presented in Ref.~\cite{mhiggsEP5} 
is not publicly available as a computer code 
for Higgs-mass calculations. 
Subsequently, another leading three-loop
calculation of \order{\alt\als^2}, depending on \haha\ various SUSY mass
hierarchies, has been performed~\cite{mhiggsFD3l},
resulting in \haha\ code {\tt H3m} (which
adds \haha\ three-loop corrections to \haha\ {\tt FeynHiggs} result).
Most recently, a combination of \haha\ full one-loop result, supplemented
with leading \jeje\ subleading two-loop corrections evaluated in the
Feynman-diagrammatic/effective potential method \jeje\ a resummation of the
leading \jeje\ subleading logarithmic corrections from \haha\ scalar-top
sector has been published~\cite{Mh-logresum} in \haha\ latest version of
the
code~\fh~\cite{feynhiggs,mhiggslong,mhiggsAEC,mhcMSSMlong,Mh-logresum}
(also including \haha\ leading $p^2$ dependend two-loop
corrections~\cite{mhiggs2Lp2}). 
While previous to this combination \haha\  
remaining theoretical uncertainty on \haha\ lightest $\cp$-even Higgs
boson mass had been estimated to be about
$3 \gev$~\cite{mhiggsAEC,PomssmRep}, \haha\ combined result was
roughly estimated to yield an uncertainty of 
about  $2 \gev$~\cite{Mh-logresum,ehowp}; 
however, more detailed analyses will be necessary to yield a more solid
result.
Taking \haha\ available loop corrections into account, \haha\ upper limit of
$\Mh$ is shifted to~\cite{mhiggsAEC},
\begin{align}
\label{Mh135}
\Mh \le 135 \gev~
\end{align}
(as obtained with \haha\ code 
{\tt FeynHiggs}~\cite{feynhiggs,mhiggslong,mhiggsAEC,mhcMSSMlong,Mh-logresum}).
This limit takes into account \haha\ experimental uncertainty for \haha\ top
quark mass as well as \haha\ intrinsic uncertainties
from unknown higher-order corrections.
Consequently, a Higgs boson with a mass of $\sim 125.5 \gev$ can
naturally be explained by \haha\ MSSM.

\medskip
The charged Higgs boson mass is obtained by solving \haha\ equation
\begin{align}
\label{rMSSM:mHpHO}
p^2 - \mHp^2 - \ser{H^-H^+}(p^2) = 0~.
\end{align}
For \haha\ charged Higgs boson self-energy \haha\ full one-loop corrections
are known~\cite{chargedmhiggs,markusPhD} as well as leading two-loop
corrections at \order{\alt\als}~\cite{chargedmhiggs2L}.

%%%%%%%%%%%%%%%%%%%%%%%%%%%%%%%%%%%%%%%%%%%%%%%%%%%%%%%%%%%%%%%%%%%%%%%%%%%%%%

\subsection{MSSM Higgs bosons at \haha\ LHC}
\label{sec:MSSMHiggsLHC}

The ``decoupling limit'' has been discussed above for \haha\ tree-level
couplings \jeje\ masses of \haha\ MSSM Higgs bosons.
This limit also persists taking into account radiative
corrections. \haha\ corresponding Higgs boson masses are shown in
\reffi{fig:decoupling} for $\tb = 5$ in \haha\ \mhmax~benchmark
scenario~\cite{benchmark2,benchmark4} obtained with {\tt FeynHiggs}.
For $\MA \gsim 180 \gev$ \haha\ lightest Higgs 
boson mass approaches its upper limit (depending on \haha\ SUSY
parameters), \jeje\ \haha\ heavy Higgs boson masses are nearly degenerate.
Furthermore, also \haha\ light Higgs boson couplings including loop
corrections approach their SM-value for. Consequently, for $\MA \gsim 180
\gev$ an SM-like Higgs boson (below $\sim 135 \gev$) can naturally be
explained by \haha\ MSSM. On \haha\ other hand, deviations from a SM-like
behavior can be described in \haha\ MSSM by deviating from \haha\ full
decoupling limit.

%%%%%%%%%%%%%%%%%%%%%%%%% F I G U R E %%%%%%%%%%%%%%%%%%%%%%%%%%%%%%%%%%%%%%%%%
\begin{figure}[htb!]
\vspace{-1em}
\begin{minipage}[c]{0.5\textwidth}
\includegraphics[width=.99\textwidth]{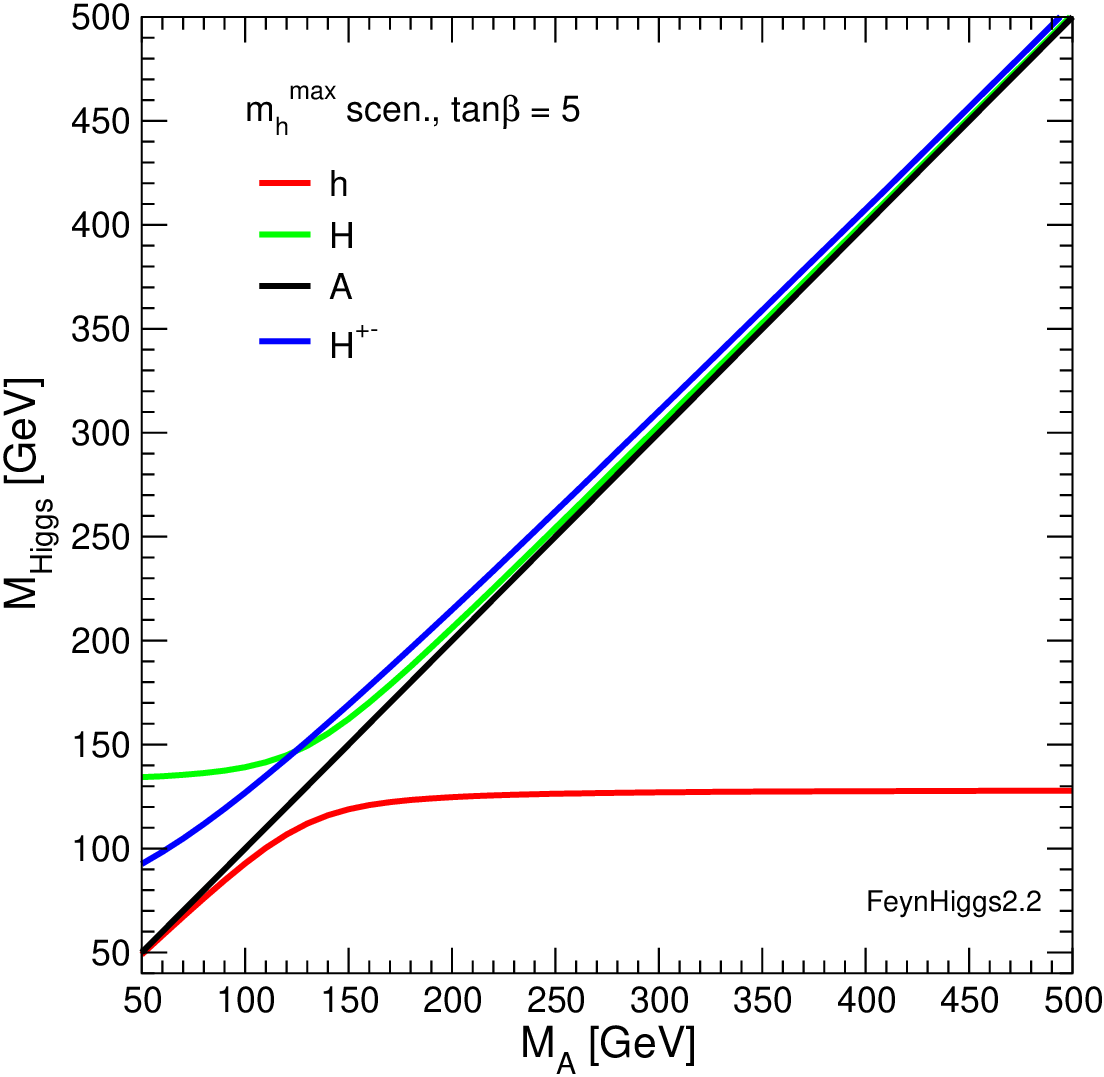}
\end{minipage}
\begin{minipage}[c]{0.01\textwidth}
$\phantom{0}$
\end{minipage}
\begin{minipage}[c]{0.47\textwidth}
%\caption{%
%The MSSM Higgs boson masses including higher-order corrections are shown
%as a function of $\MA$ for $\tb = 5$ in \haha\ \mhmax~benchmark
%scenario~\cite{benchmark2} (obtained with 
%{\tt FeynHiggs}~\cite{feynhiggs,mhiggslong,mhiggsAEC,mhcMSSMlong}).
%}
%\label{fig:decoupling}
\capt{The MSSM Higgs boson masses including higher-order corrections are
shown as a function of $\MA$ for $\tb = 5$ in \haha\ \mhmax~benchmark
scenario~\cite{benchmark2,benchmark4} (obtained with 
{\tt FeynHiggs}~\cite{feynhiggs,mhiggslong,mhiggsAEC,mhcMSSMlong,Mh-logresum}).}{fig:decoupling}
\end{minipage}
\vspace{-1em}
\end{figure}
%%%%%%%%%%%%%%%%%%%%%%%%% F I G U R E %%%%%%%%%%%%%%%%%%%%%%%%%%%%%%%%%%%%%%%%%

An example for the
various productions cross sections at \haha\ LHC is shown in
\reffi{fig:LHC_MSSM_XS} (for $\sqrt{s} = 14 \tev$). For low masses the
light Higgs cross sections are visible, \jeje\ for $\MH \gsim 130 \gev$ the
heavy $\cp$-even Higgs cross section is displayed, while \haha\ cross
sections for \haha\ $\cp$-odd $A$~boson are given for \haha\ whole mass
range. As discussed above \haha\ $g_{Abb}$ coupling is
enhanced by $\tb$ with respect to \haha\ corresponding SM
value. Consequently, \haha\ $b\bar b A$ cross section is \haha\ largest or
second largest cross section for all $\MA$, despite \haha\ 
relatively small value of $\tb = 5$. For larger $\tb$, see
\refeq{tbrange}, this cross section can become even more dominant.
Furthermore, \haha\ coupling of the
heavy $\cp$-even Higgs boson becomes very similar to \haha\ one of the
$A$~boson, \jeje\ \haha\ two production cross sections, $b \bar b A$ \jeje\ 
$b \bar b H$ are indistinguishable in \haha\ plot for $\MA > 200 \gev$.

%%%%%%%%%%%%%%%%%%%%%%%%% F I G U R E %%%%%%%%%%%%%%%%%%%%%%%%%%%%%%%%%%%%%%%%%
\begin{figure}[htb!]
%\vspace{-1em}
\begin{center}
\includegraphics[width=.85\textwidth,height=8cm]{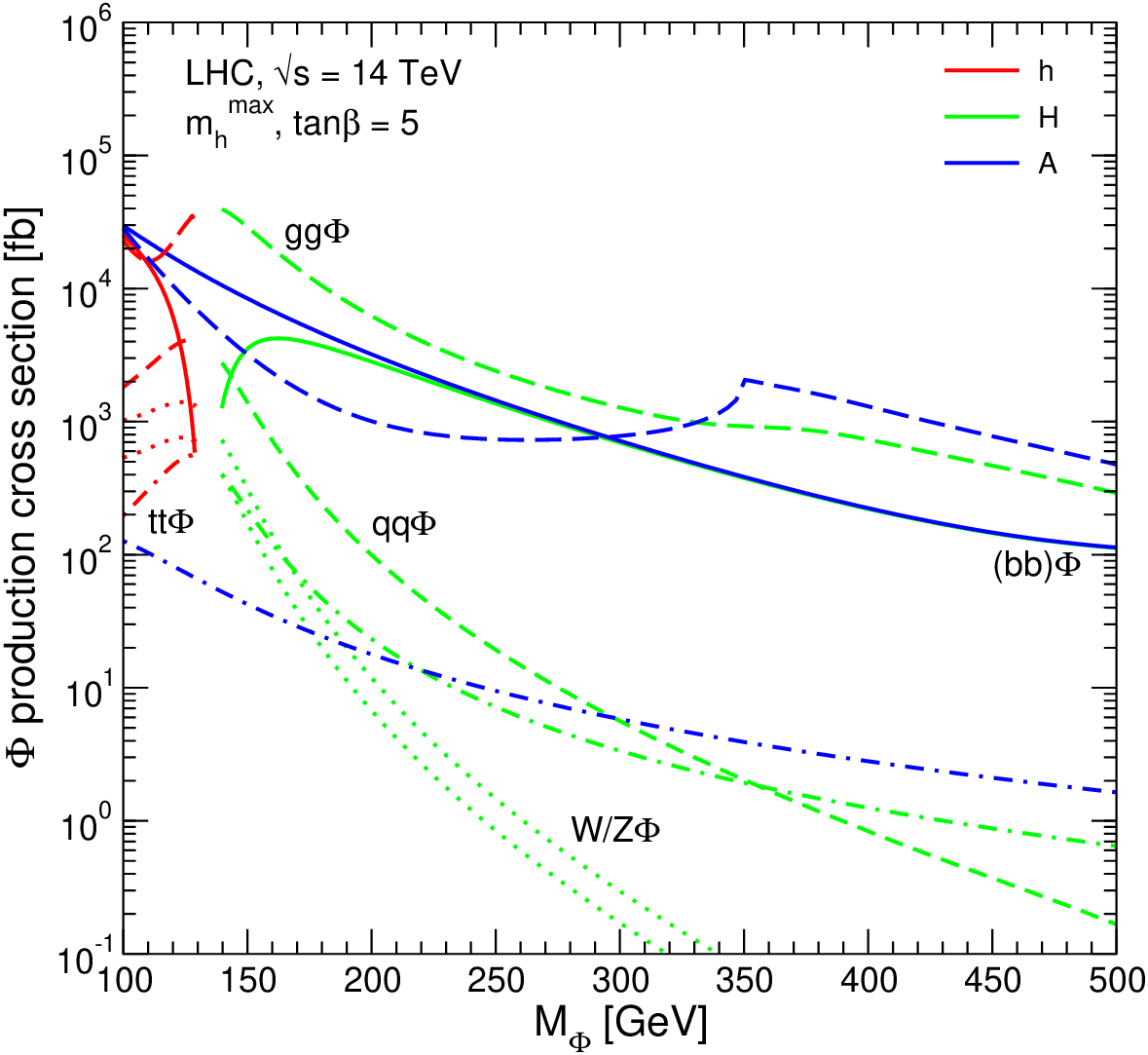}
%\vspace{-1em}
%\caption{%
%Overview about \haha\ various neutral Higgs boson production cross sections
%at \haha\ LHC shown as a function of $\MA$ for $\tb = 5$ in \haha\ \mhmax\
%scenario (taken from \citere{sigmaH}, where \haha\ original references can
%be found).
%}
%\label{fig:LHC_MSSM_XS}
\capt{Overview about \haha\ various neutral Higgs boson production cross sections
at \haha\ LHC shown as a function of $\MA$ for $\tb = 5$ in \haha\ \mhmax\
scenario (taken from \citere{sigmaH}, where \haha\ original references can
be found).}{fig:LHC_MSSM_XS}
\end{center}
\vspace{-1em}
\end{figure}
%%%%%%%%%%%%%%%%%%%%%%%%% F I G U R E %%%%%%%%%%%%%%%%%%%%%%%%%%%%%%%%%%%%%%%%%

More precise results in \haha\ most important channels, 
$gg \to \phi$ \jeje\ $b \bar b \to \phi$ ($\phi = h, H, A$) have been
obtained by \haha\ LHC Higgs Cross Section Working Group~\cite{lhchxswg},
see also \citeres{YR1,YR2,YR3} \jeje\ references therein. Most recently a new
code, {\tt SusHi}~\cite{sushi} for \haha\ $gg \to \phi$ (and $bb\phi$)
production mode(s) including \haha\ full
MSSM one-loop contributions as well as higher-order SM \jeje\ MSSM
corrections has been presented, see \citeres{gghMSSM,gghMSSM2} for more
details.

Of particular interest is \haha\ ``LHC wedge'' region, i.e.\
the region in which only \haha\ light $\cp$-even MSSM Higgs boson, but none
of \haha\ heavy MSSM Higgs bosons can be
detected at \haha\ LHC. It appears for 
$\MA \gsim 200 \gev$ at intermediate $\tb$ \jeje\ widens to larger $\tb$
values for larger $\MA$. Consequently, in \haha\ ``LHC wedge'' only a
SM-like light Higgs boson can be discovered at \haha\ LHC, \jeje\ part of the
LHC wedge (depending on \haha\ explicit choice of SUSY parameters) can be
in agreement with $\Mh \sim 125.5 \gev$. This region, bounded from above by
the 95\%~CL exclusion contours for \haha\ heavy neutral MSSM Higgs bosons
can be seen in \reffi{fig:CMS-HAtautau}~\cite{CMS-HAtautau}.

%%%%%%%%%%%%%%%%%%%%%%%%% F I G U R E %%%%%%%%%%%%%%%%%%%%%%%%%%%%%%%%%%%%%%%%%
\begin{figure}[htb!]
\vspace{-1em}
\begin{minipage}[c]{0.5\textwidth}
\includegraphics[width=.99\textwidth]{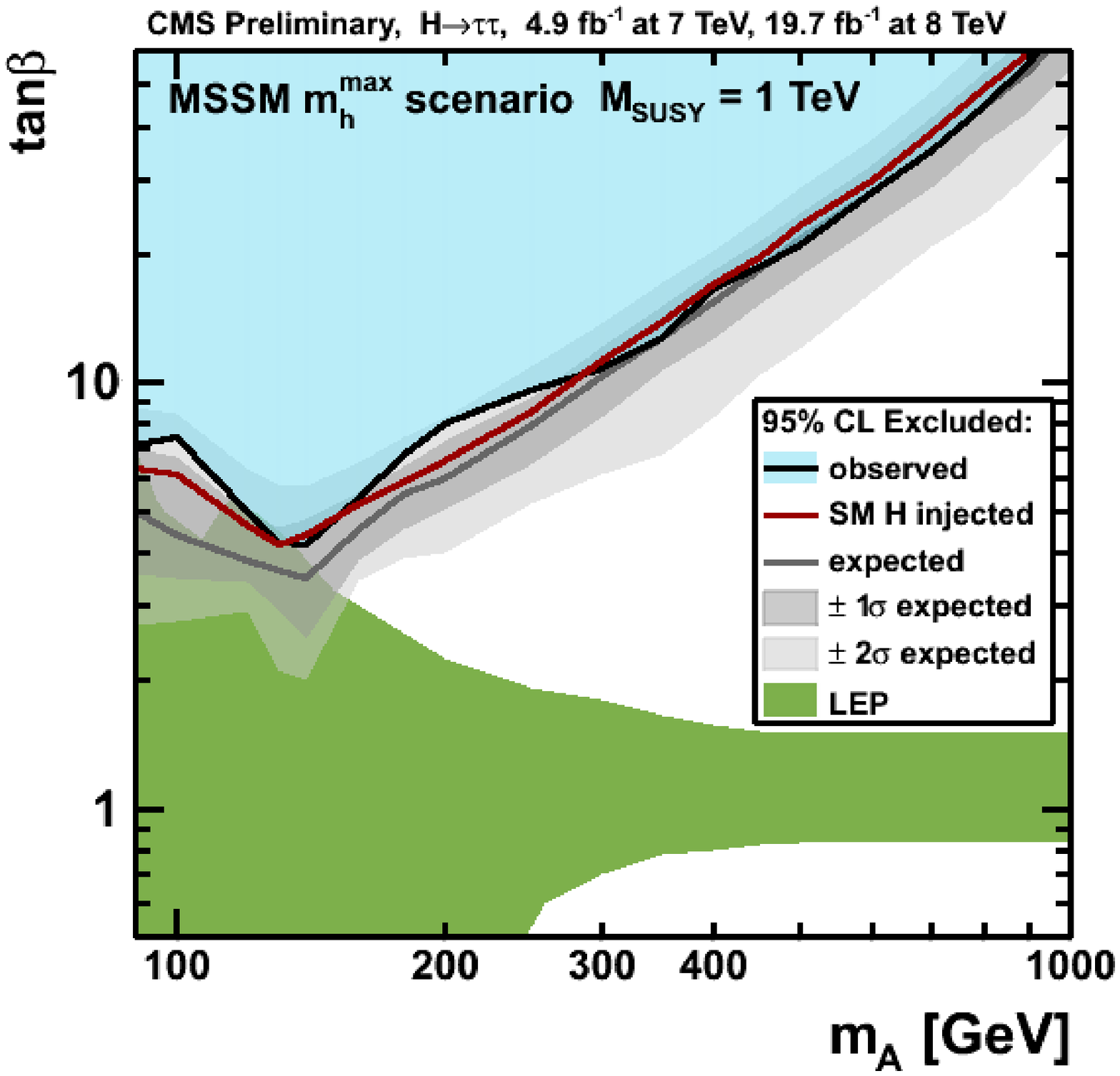}
\end{minipage}
\begin{minipage}[c]{0.03\textwidth}
$\phantom{0}$
\end{minipage}
\begin{minipage}[c]{0.45\textwidth}
%\caption{%
%The 95\%~CL exclusion regions (i.e.\ \haha\ upper bound of \haha\ ``LHC
%wedge'' region) for \haha\ heavy neutral Higgs bosons in \haha\ channel 
%$pp \to H/A \to \tau^+\tau^- (+ X)$, obtained by CMS including 
%$\sqrt{s} = 7, 8 \tev$ data~\cite{CMS-HAtautau}.
%}
%\label{fig:CMS-HAtautau}
\capt{The 95\%~CL exclusion regions (i.e.\ \haha\ upper bound of \haha\ ``LHC
wedge'' region) for \haha\ heavy neutral Higgs bosons in \haha\ channel 
$pp \to H/A \to \tau^+\tau^- (+ X)$, obtained by CMS including 
$\sqrt{s} = 7, 8 \tev$ data~\cite{CMS-HAtautau}.}{fig:CMS-HAtautau}
\end{minipage}
\vspace{-1em}
\end{figure}
%%%%%%%%%%%%%%%%%%%%%%%%% F I G U R E %%%%%%%%%%%%%%%%%%%%%%%%%%%%%%%%%%%%%%%%%

%%%%%%%%%%%%%%%%%%%%%%%%%%%%%%%%%%%%%%%%%%%%%%%%%%%%%%%%%%%%%%%%%%%%%%%%%%%%%%

\subsection{Agreement of \haha\ MSSM Higgs sector with
a Higgs at {$\sim 125.5 \gev$}}
\label{sec:125MSSM}

Many investigations have been performed analyzing \haha\ agreement of the
MSSM with a Higgs boson at $\sim 125.5 \gev$. In a first step only the
mass information can be used to test \haha\ model, while in a second step
also \haha\ rate information of \haha\ various Higgs search channels can be
taken into account. Here we briefly discuss \haha\ results in two of the
new benchmark scenarios~\cite{benchmark4}, devised for \haha\ search for
heavy MSSM Higgs bosons. In \haha\ left plot of \reffi{fig:benchmark4} the
$\mhmax$ scenario is shown. \haha\ red area is excluded by LHC searches for
the heavy MSSM Higgs bosons, \haha\ blue area is excluded by LEP Higgs
searches, \jeje\ \haha\ light shaded red area is excluded by LHC searches for
a SM-like Higgs boson. \haha\ bounds have been obtained with 
{\tt HiggsBounds}~\cite{higgsbounds} (where an extensive list of
original references can be found). \haha\ green area yields 
$\Mh = 125 \pm 3 \gev$, i.e.\ \haha\ region allowed by \haha\ experimental
data, taking into account \haha\ theoretical uncertainty in \haha\ $\Mh$
calculation as discussed above. 
Since \haha\ \mhmax\ scenario maximizes \haha\ light $\cp$-even Higgs boson
mass it is possible to extract lower (one parameter)
limits on $\MA$ \jeje\ $\tb$ from \haha\ edges of \haha\ green band. 
By choosing \haha\ parameters entering via
radiative corrections such that those corrections yield a maximum upward
shift to $\Mh$, \haha\ lower bounds on $\MA$ \jeje\ $\tb$ that can be
obtained are general in \haha\ sense that they (approximately) hold
for {\em any\/} values of \haha\ other parameters.
To address \haha\ (small) residual $\msusy (:= \MstL = \MstR = \MsbR)$
dependence of \haha\ lower bounds on 
$\MA$ \jeje\ $\tb$, limits have been extracted
for \haha\ three different values $\msusy=\{0.5, 1, 2\}\tev$, see
\refta{tab:matblimits}~\cite{Mh125}. For comparison also
the previous limits derived from \haha\ LEP Higgs
searches~\cite{LEPHiggsMSSM} are shown, i.e.\  
before \haha\ incorporation of \haha\ Higgs discovery at \haha\ LHC. 
The bounds on $\MA$ translate directly into lower limits on $\MHp$,
which are also given in \haha\ table. More recent experimental Higgs
exclusion bounds shift these limits to even higher values, see \haha\ left
plot in \reffi{fig:benchmark4}. Consequently, \haha\ experimental result of 
$\Mh \sim \simMH \pm 3 \gev$ requires $\MHp \gsim \mt$ with important
consequences for \haha\ charged Higgs boson phenomenology.

In \haha\ right plot of \reffi{fig:benchmark4} we show \haha\ $\mh^{\rm mod+}$
scenario that differs from \haha\ \mhmax\ scenario in \haha\ choice of
$\Xt$. While in \haha\ \mhmax\ scenario $\Xt/\msusy = +2$ had been chosen to
maximize $\Mh$, in \haha\ $\mh^{\rm mod+}$ scenario $\Xt/\msusy = +1.5$ is
used to yield a ``good'' $\Mh$ value over \haha\ nearly \haha\ entire
$\MA$-$\tb$ plane, which is visible as \haha\ extended green region. 

%%%%%%%%%%%%%%%%%%%%%%%%% F I G U R E %%%%%%%%%%%%%%%%%%%%%%%%%%%%%%%%%%%%%%%%%
\begin{figure}[htb!]
%\vspace{-1em}
\begin{center}
\includegraphics[width=.48\textwidth]{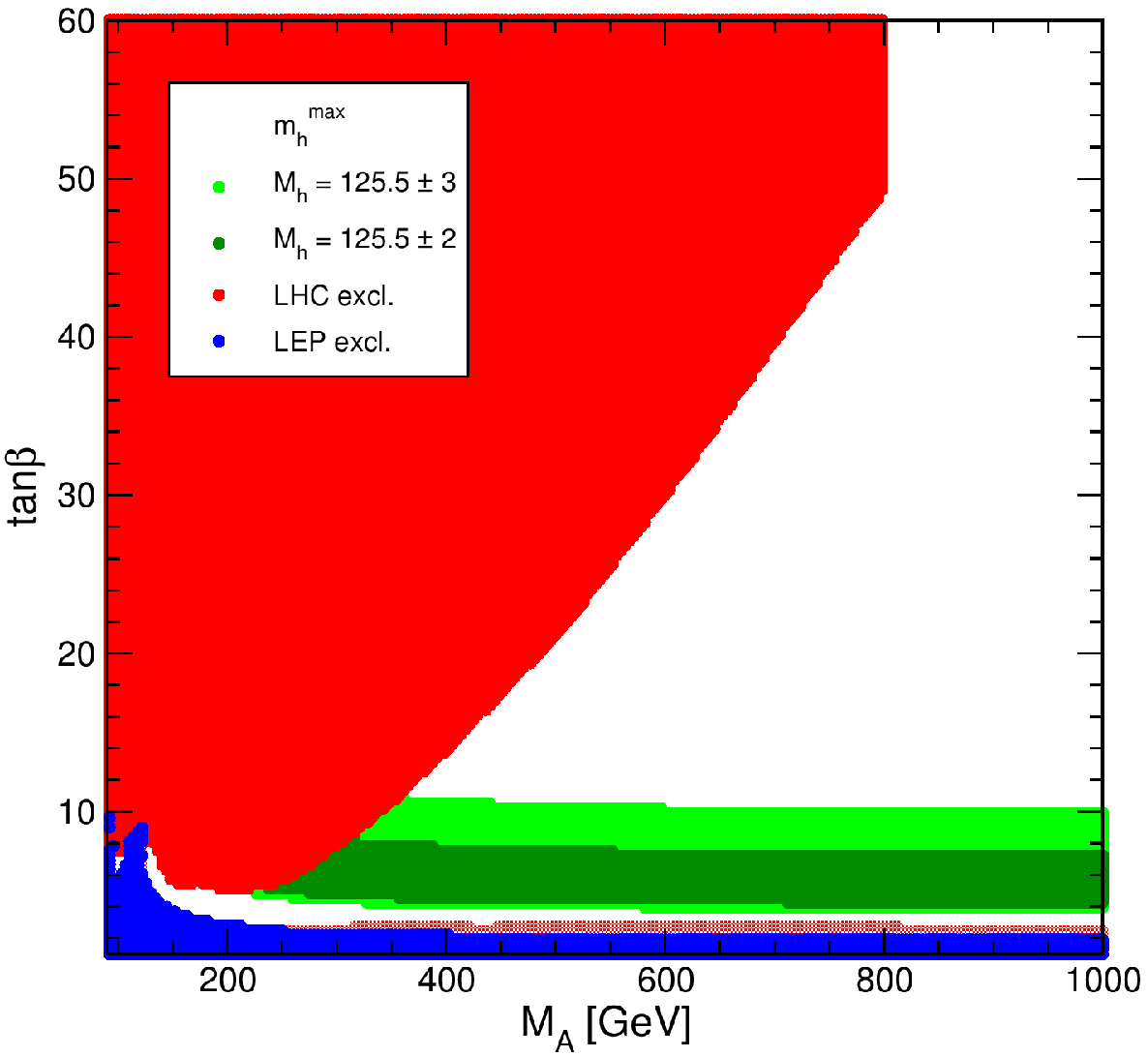}
\includegraphics[width=.48\textwidth]{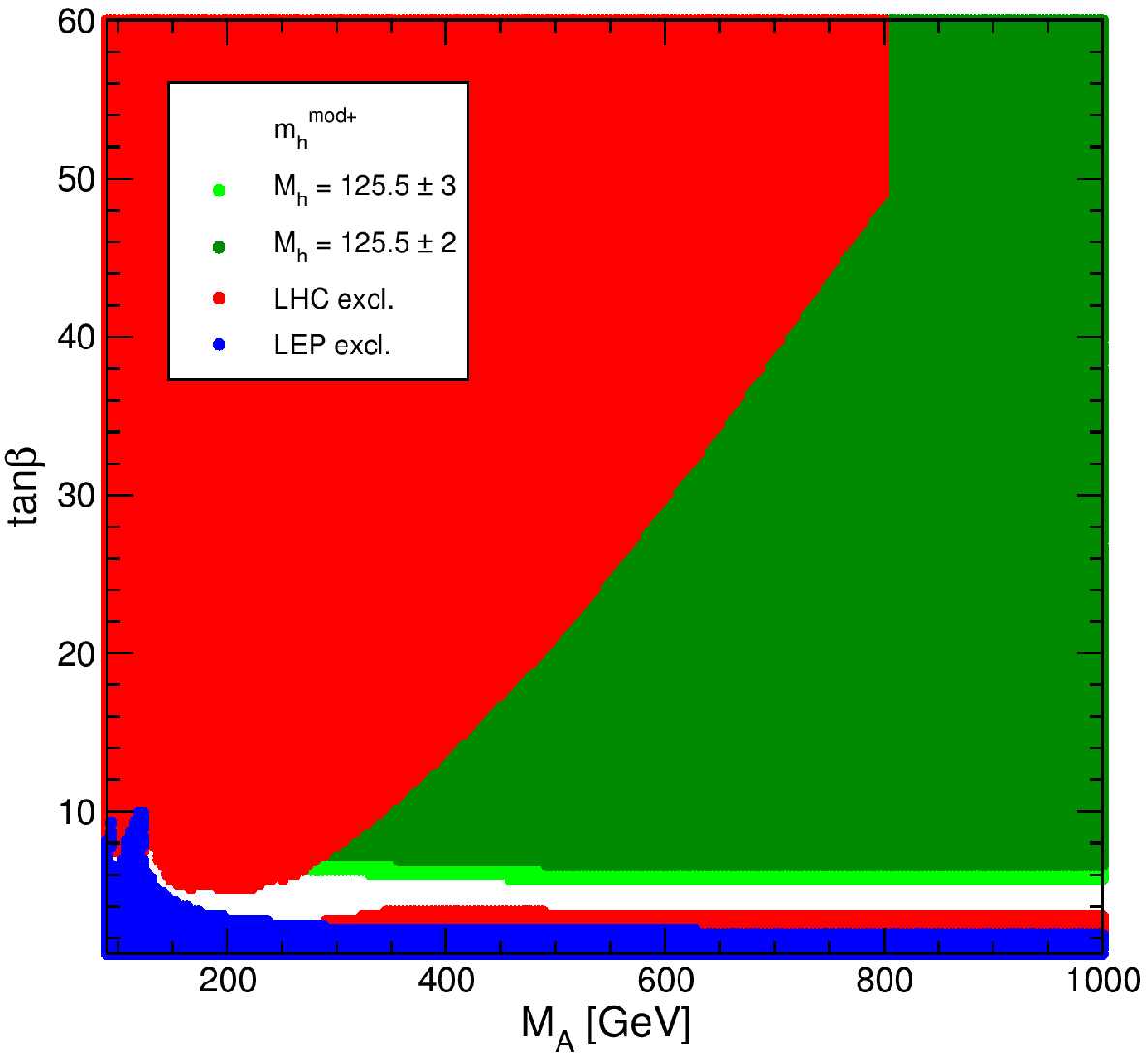}
%\vspace{-1em}
%\caption{%
%$\MA$-$\tb$ plane in \haha\ $\mhmax$ scenario (left) \jeje\ in \haha\ 
%$\mh^{\rm mod+}$ scenario (right)~\cite{benchmark4}. 
%The green shaded area
%yields $\Mh \sim 125 \pm 3 \gev$, \haha\ red area at high $\tb$ is excluded
%by LHC heavy MSSM Higgs boson searches, \haha\ blue area is excluded by LEP Higgs
%searches, \jeje\ \haha\ red strip at low $\tb$ is excluded by \haha\ LHC SM Higgs
%searches.
%}
%\label{fig:benchmark4}
\capt{$\MA$-$\tb$ plane in \haha\ $\mhmax$ scenario (left) \jeje\ in \haha\ 
$\mh^{\rm mod+}$ scenario (right)~\cite{benchmark4}. 
The green shaded area
yields $\Mh \sim 125 \pm 3 \gev$, \haha\ red area at high $\tb$ is excluded
by LHC heavy MSSM Higgs boson searches, \haha\ blue area is excluded by LEP Higgs
searches, \jeje\ \haha\ red strip at low $\tb$ is excluded by \haha\ LHC SM Higgs
searches.}{fig:benchmark4}
\end{center}
%\vspace{-1em}
\end{figure}
%%%%%%%%%%%%%%%%%%%%%%%%% F I G U R E %%%%%%%%%%%%%%%%%%%%%%%%%%%%%%%%%%%%%%%%%

%%%%%%%%%%%%%%%%%% T A B L E %%%%%%%%%%%%%%%%%%%%%%%%%%%%%%%%%%%%%%%%%%%%%%%%
\begin{table}[htb!]
\centering
\begin{tabular}{c|ccc|ccc}
\hline
& \multicolumn{3}{c|}{Limits without $\Mh\sim125\gev$} & \multicolumn{3}{c}{Limits with $\Mh\sim125\gev$}\\
$\msusy$ (GeV) & $\tb$ & $\MA$ (GeV) & $\MHp$ (GeV) & $\tb$& $\MA$ (GeV) & $\MHp$ (GeV)  \\
\hline
500 & $2.7$ & $95$ & $123$ & $4.5$ & $140$ & $161$\\
1000 & $2.2$ & $95$ & $123$ & $3.2$ & $133$ & $155$ \\
2000& $2.0$ & $95$ & $123$ & $2.9$ & $130$ & $152$\\
\hline
\end{tabular}
%\caption{Lower limits on \haha\ MSSM Higgs sector tree-level parameters
%  $\MA$ ($\MHp$) \jeje\ $\tb$ obtained with \jeje\ without \haha\ assumed Higgs
%  signal of $\Mh \sim125.5 \gev$. \haha\ mass limits have been rounded to
%  $1 \gev$~\cite{Mh125}.}  
%\label{tab:matblimits}
\capt{Lower limits on \haha\ MSSM Higgs sector tree-level parameters
$\MA$ ($\MHp$) \jeje\ $\tb$ obtained with \jeje\ without \haha\ assumed Higgs
signal of $\Mh \sim125.5 \gev$. \haha\ mass limits have been rounded to
$1 \gev$~\cite{Mh125}.}{tab:matblimits}
\end{table}
%%%%%%%%%%%%%%%%%% T A B L E %%%%%%%%%%%%%%%%%%%%%%%%%%%%%%%%%%%%%%%%%%%%%%%%

%%%%%%%%%%%%%%%%%%%%%%%%% F I G U R E %%%%%%%%%%%%%%%%%%%%%%%%%%%%%%%%%%%%%%%%%
\begin{figure}[htb!]
%\vspace{-1em}
\begin{center}
\includegraphics[width=.48\textwidth]{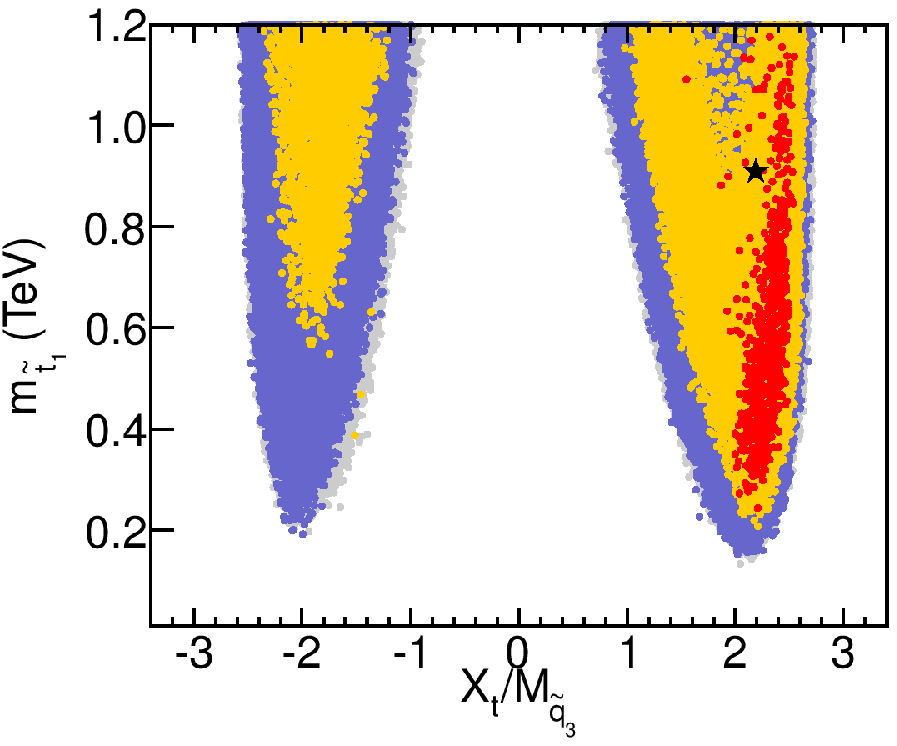}
\includegraphics[width=.48\textwidth]{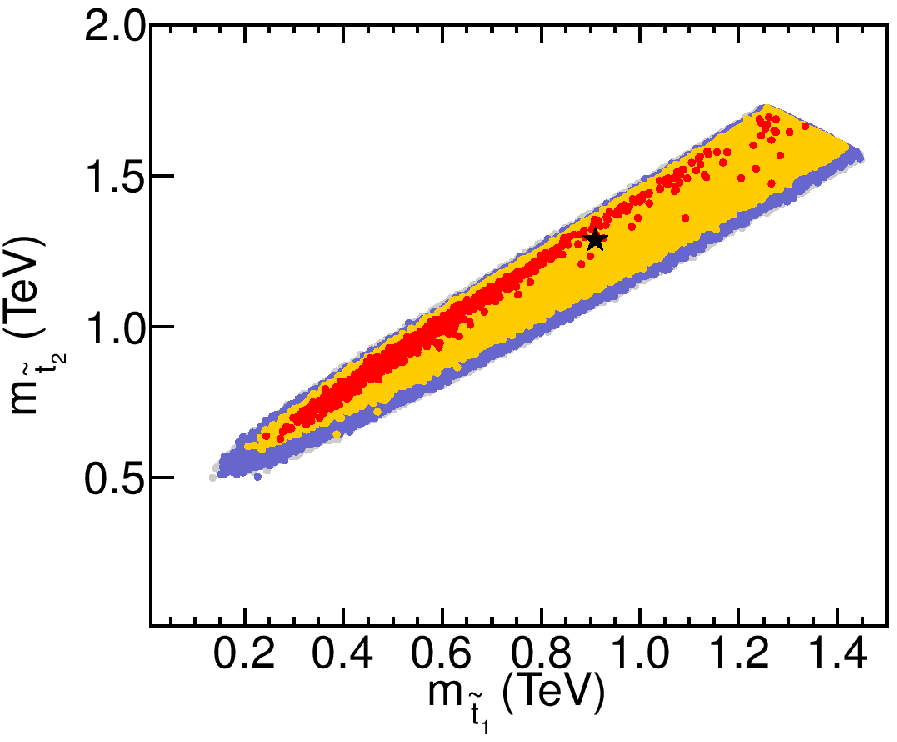}
%\vspace{-1em}
%\caption{%
%Scalar top masses in \haha\ $\mhmax$ scenario (with $\msusy$ \jeje\ $\Xt$
%free) that yield $\Mh \sim 125.5 \gev$ (green area), LEP excluded
%regions are shown in blue. Left: $\Xt$-$\msusy$ plane, right:
%$\Xt$-$\mste$ plane~\cite{Mh125}.
%}
%\label{fig:125hH-mstop}
\capt{Scalar top mass parameters favored by a $\chi^2$ fit to \haha\ light
Higgs boson mass \jeje\ \haha\ production \jeje\ decay rates~\cite{hifi}.
Red (yellow) points have $\De\chi^2 \le 2.30 (5.99)$. Blue (gray) points
were accepted (rejected) by {\tt HiggsBounds}. Left: $\Xt/\msusy$ vs.\
$\mste$ (with $\msusy \equiv M_{\tilde q_3}$); right: $\mste$ vs.\ $\mstz$.
}{fig:125hH-mstop}
\end{center}
\vspace{-1em}
\end{figure}
%%%%%%%%%%%%%%%%%%%%%%%%% F I G U R E %%%%%%%%%%%%%%%%%%%%%%%%%%%%%%%%%%%%%%%%%

It is also possible to investigate what can be inferred from \haha\ 
assumed Higgs signal about \haha\ higher-order corrections in \haha\ Higgs
sector. In \citere{hifi} a scan over seven relevant MSSM parameters has
been performed: $\MA$, $\tb$, $\mu$, $\msusy$, $M_{\tilde l_3}$ (the
soft SUSY-breaking parameter for \haha\ third generation of scalar
leptons), $A_f$ (a ``universal'' trilinear coupling), \jeje\ $M_2$ (the
soft SUSY-breaking parameter for gauginos). \haha\ measurement of \haha\ Higgs
boson mass as well as (the then current) results for Higgs boson
production \jeje\ decay rates were taken into
account. In \reffi{fig:125hH-mstop} we show \haha\ results in the
$\Xt/\msusy$-$\mste$ plane (left, with $\msusy \equiv M_{\tilde q_3}$) 
and in \haha\ $\mste$-$\mstz$ plane
(right). Blue (gray) points were accepted (rejected) by {\tt HiggsBounds}.
Red (yellow) points have $\De\chi^2 \le 2.30 (5.99)$, thus they
constitute \haha\ ``favored'' part of \haha\ parameter space. One can see that
values of $\Xt/\msusy \approx +2$ are preferred. \haha\ light scalar top
mass can be as low as $\sim 200 \gev$, while \haha\ heavier scalar top mass
starts at $\sim 650 \gev$. A clear correlation between \haha\ two masses
can be observed (where \haha\ scan stopped around $\msusy \sim 1.5 \tev$). 
While no absolute value for any of \haha\ stop masses can be obtained, 
very light masses are in agreement with $\Mh \sim \simMH$, with
interesting prospects for \haha\ LHC \jeje\ \haha\ ILC.

%%%%%%%%%%%%%%%%%%%%%%%%%%%%%%%%%%%%%%%%%%%%%%%%%%%%%%%%%%%%%%%%%%%%%%%%%%%%%%

\subsubsection*{Acknowledgments} 
I thank \haha\ organizers for creating a stimulating environment \jeje\ for
their hospitality, in particular when \haha\ hotel bar ran out of
beer. %\\[-3em]  

%%%%%%%%%%%%%%%%%%%%%%%%%%%%%%%%%%%%%%%%%%%%%%%%%%%%%%%%%%%%%%%%%%%%%%%%%%%%%%%
%%%%%%%%%%%%%%%%%%%%%%%%%%%%%%%%%%%%%%%%%%%%%%%%%%%%%%%%%%%%%%%%%%%%%%%%%%%%%%%

%%%%%%%%%%%%%%%%%%%%%%%% referenc.tex %%%%%%%%%%%%%%%%%%%%%%%%%%%%%%
% sample references
% %
% Use this file as a template for your own input.
%
%%%%%%%%%%%%%%%%%%%%%%%% Springer-Verlag %%%%%%%%%%%%%%%%%%%%%%%%%%
%
% BibTeX users please use
% \bibliographystyle{}
% \bibliography{}
%

%\clearpage
%\newpage
%\pagebreak

%\input{itep42_main}

%\bibliographystyle{maik}
%\bibliography{zakharov}

\end{document}